\documentclass[%
nofootinbib,
 amsmath,amssymb,
 aps,
algorithm,
]{revtex4-2}

\usepackage{graphicx}
\usepackage{dcolumn}
\usepackage{bm}
\usepackage{xcolor}

\usepackage{algorithm}
\usepackage{algpseudocode}	
\usepackage{booktabs} 
\usepackage{multirow}
\usepackage{longtable}
\usepackage{tabularx}

\usepackage{kotex} 

\newcommand{\best}{\textsc{CMB-BEST}}    
\newcommand{\bestcode}{\texttt{cmbbest}}    
\newcommand{\tetraquad}{\texttt{tetraquad}}    

\newcommand{\vv}{\mathbf}			
\newcommand{\pluseq}{\mathrel{+}=}	

\newcommand{\ells}{{\ell_1 \ell_2 \ell_3}}

\newcommand{\As}{{A_\mathrm{s}}}
\newcommand{\ns}{{n_\mathrm{s}}}

\newcommand{\fNL}{f_\mathrm{NL}}
\newcommand{\kmin}{k_\mathrm{min}}
\newcommand{\kmax}{k_\mathrm{max}}
\newcommand{\kratio}{k_\mathrm{r}}

\newcommand{\tetra}{\mathcal{V}_\mathrm{T}}     
\newcommand{\tetrasix}{\mathcal{V}_{\mathrm{T}/6}}     

\newcolumntype{P}[1]{>{\centering\arraybackslash}p{#1}}     

\begin{document}


\title{High-resolution CMB bispectrum estimator with flexible modal basis}

\author{Wuhyun Sohn}
 \email{wuhyun@kasi.re.kr}
\affiliation{%
Korea Astronomy and Space Science Institute, Daejeon 34055, South Korea
}
\author{James R. Fergusson}%
 \email{jf334@cam.ac.uk}
\affiliation{%
Centre for Theoretical Cosmology, Department of Applied Mathematics and Theoretical Physics, University of Cambridge, Cambridge CB3 0WA, United Kingdom
}

\author{E. P. S. Shellard}%
 \email{epss@damtp.cam.ac.uk}
\affiliation{%
Centre for Theoretical Cosmology, Department of Applied Mathematics and Theoretical Physics, University of Cambridge, Cambridge CB3 0WA, United Kingdom
}

\date{\today}
\hspace{10pt}

\begin{abstract}

We present a new independent pipeline for the CMB bispectrum estimation of primordial non-Gaussianity and release a public code for constraining bispectrum shapes of interest based on the Planck 2018 temperature and polarization data. The estimator combines the strengths of the conventional KSW and Modal estimators at the cost of increased computational complexity, which has been made manageable through intensive algorithmic and implementation optimization. We also detail some methodological advances in numerical integration over a tetrapyd\textemdash domain where the bispectrum is defined on\textemdash via new quadrature rules. The pipeline has been validated both internally and against Planck. As a proof-of-concept example, we constrain some highly oscillatory models that were out of reach in conventional analyses using a targeted basis with a fixed oscillation frequency, and no significant evidence for primordial non-Gaussianity of these shapes is found. The methodology and code developed in this work will be directly applicable to future surveys where we expect a notable boost in sensitivity.

\end{abstract}

\maketitle


\section{Introduction}

To what degree the primordial perturbations are, if at all, non-Gaussian, is a key question of cosmology with many implications for early universe physics \cite{Meerburg2019,Chen2010review,Komatsu2010,Liguori2010,Desjacques2010PNGLSS,RenauxPetel2015PNGreview,Bartolo2002NGinflation,Maldacena2003,Bartolo2004}. Most theoretically well-motivated models predict various amplitudes and shapes of primordial non-Gaussianity (PNG, see e.g. the reviews of \cite{Meerburg2019,Chen2010review,Komatsu2010,Liguori2010}), while the simplest single-field slow-roll inflation predicts an undetectably small PNG \cite{Maldacena2003}. In a non-Gaussian universe, statistics beyond the two-point function are required to capture the full statistical information. The primordial bispectrum, or the harmonic counterpart of the three-point function, is the next leading-order measure of PNG. All cosmological datasets at present are consistent with vanishing PNG \cite{PlanckCollaboration2013,PlanckCollaboration2015,PlanckCollaboration2018,WMAP2013maps,Mueller2022eBOSSPNG,Cabass2022BOSSPNG}. Among them, Planck's CMB bispectrum analysis has placed the most stringent constraints on PNG to date \cite{PlanckCollaboration2018}. 

The CMB bispectrum estimation of PNG, often parametrized by its amplitude $\fNL$, is a computationally challenging task. A na\"ive computation is prohibitively expensive, so all existing implementations utilize some simplifying techniques. The KSW (or KSW-like) estimators \cite{Komatsu2005,Creminelli2006limits,Yadav2007,Senatore2010orthogonal,Smith2011} and skew-$C_\ell$ statistics \cite{Munshi2010skewCl} utilize separable (or factorizable) bispectrum templates which dramatically simplifies the integrals involved. The Modal estimator \cite{Fergusson2010general,Fergusson2012,Fergusson2014} uses separable modal expansions of the primordial and the late-time bispectrum. The Binned bispectrum estimator \cite{Bucher2010,Bucher2016} compresses the data by binning the bispectrum in harmonic space with minimal loss of optimality.

A great variety of physically well-motivated models have been tested by the Planck collaboration \cite{PlanckCollaboration2013,PlanckCollaboration2015,PlanckCollaboration2018} using these estimators. Among them are models with oscillations in the bispectrum induced by, e.g., features in the inflationary potential \cite{Starobinsky1992,Adams2001,Covi2006WMAPfeature,Chen2007,Chen2008featureNG,Adshead2011,Adshead2012,Hazra2014}, a transient reduction in the speed of sound \cite{Achucarro2011,Nakashima2011soundspeed,Park2012soundspeed,Achucarro2014transient,Achucarro2014}, multi-field dynamics \cite{Gao2012sharpturn,Gao2013sharpturn,Noumi2013effective,Noumi2013sharpturn}, or resonances arising in axion monodromy models \cite{Silverstein2008monodromy,McAllister2010monodromy,Flauger2010monodromy,Berg2010monodromyTwoField,Flauger2011ResonantNG,Aich2013oscillation,Behbahani2012resonantNG} (see e.g., \cite{Chen2010review,Chluba2015,Slosar2019review,Achucarro2022Inflation} for reviews). Many of these models predict some enveloped oscillations in the primordial power spectrum and/or bispectrum that are linearly or logarithmically spaced. Previous works have placed constraints on these models using the CMB power spectrum \cite{Martin2004,Pahud2009oscillations,Hazra2010WMAPfeatures,Dvorkin2011WMAPenvelopeFeatures,Benetti2011WMAPfeatures,Meerburg2012,Chen2012standardClock,Benetti2013updates,Hazra2013MRL,Meerburg2014a,Meerburg2014b,Meerburg2014,PlanckCollaboration2013power,Chen2014standardClock,PlanckCollaboration2015power,PlanckCollaboration2015isotropy,PlanckCollaboration2018power,Canas2021BayesianCs}, bispectrum \cite{Meerburg2010oscillations,Munchmeyer2014,Munchmeyer2015resonance,Meerburg2015excited,PlanckCollaboration2013,PlanckCollaboration2015,PlanckCollaboration2018}, both in a joint analysis \cite{Fergusson2015a,Fergusson2015b,Meerburg2016jointResonance}, and in combination with or solely from large-scale structure data \cite{Hamann2007SDSSFeatureConstraints,Hu2015,Benetti2016,Beutler2019}. However, despite its significant implications for early universe physics, constraining highly oscillatory bispectrum shapes with general (non-separable) envelopes has been out of reach using conventional methods due to computational challenges \cite{PlanckCollaboration2018}. There remains a variety of such models that are yet unconstrained or only partially constrained

Furthermore, to the authors' knowledge, there currently is no publicly available code that allows one to get Planck CMB constraints on a general bispectrum shape of interest \footnote{We note that public codes for the KSW \cite{KSWcode} and binned bispectrum estimator \cite{BinnedBispectrumcode} are available but do not apply to a general shape.}. Having access to CMB bispectrum constraints would greatly benefit researchers in testing various models of interest. 

Motivated by these reasons, we developed an independent bispectrum estimation pipeline \best. The method generalizes the KSW estimator by allowing a flexible choice of basis functions, which is used for a Modal-like expansion of general bispectrum templates. It combines the accuracy of KSW with the broad applicability of Modal at the cost of increased computational cost. We have extensively optimized the algorithm and implementation of \best\ to make computation manageable. We have also thoroughly tested the code for self-consistency and against Planck's Modal estimator. Some new constraints on a class of highly oscillatory templates, as well as reproductions of the Planck 2018 analyses, are presented in this paper as proof-of-concept examples.

We publicly release the frontend of our code as a Python package named \texttt{cmbbest}, where users can provide general bispectrum shapes of interest and obtain Planck 2018 constraints on the corresponding $\fNL$s in a matter of seconds, or minutes for a higher-resolution basis, on a laptop. This was made possible by pre-computing the computationally expensive parts of the pipeline on super-computing clusters and providing the results as a data file. We plan on providing more basis function choices and future survey data in time.

There have been some methodological advances during the development of \best. This paper includes our novel quadrature rule for efficient numerical integration of functions over a `tetrapyd' domain where the bispectrum is defined. The method is much more efficient and accurate than the simple 3D trapezoidal rule and is expected to benefit some numerical analyses of the large-scale structure as well. The method is implemented and shared as a Python package \texttt{tetraquad}. 

Upcoming CMB experiments such as the Simons Observatory \cite{TheSimonsObservatoryCollaboration2018} and CMB-S4 \cite{Abazajian2016,Abazajian2019} are expected to dramatically improve the sensitivity in polarization measurements. The future datasets, in combination with the existing ones, will provide constraints on PNG that are almost as stringent as they can be from the CMB alone \cite{Abazajian2019}. The methodology and code developed here will be directly applicable to upcoming surveys.

This paper is organized as follows. In Section \ref{section:CMB_bispectrum_estimation}, we review the CMB bispectrum estimation procedure. Section \ref{section:high-resolution_CMB_bispectrum_estimator} details the main formalism of \best\ and various basis function choices made in this work. Section \ref{section:numerical_integration_over_tetrapyd} introduces our novel numerical method for evaluating integrals over a tetrapyd domain. Section \ref{section:results} details the public release of our code and provides some proof-of-concept examples of \best. Appendices contain some computational details and various consistency checks. The conclusion is given in Section \ref{section:conclusion}.

\section{CMB Bispectrum estimation}  \label{section:CMB_bispectrum_estimation}

In this section, we review how the CMB bispectrum can be used to study PNG. We provide a derivation of the CMB bispectrum estimator from the CMB bispectrum likelihood written analogously to \cite{Liguori2010,Becker2012wmapRunningfNL,Meerburg2016jointResonance}. We note that the core ideas behind this formalism are heavily based on the original works on the topic \cite{Heavens1998bispectrumEstimator,Komatsu2001acousticCMBBispectrum,Komatsu2005,Yadav2007,Creminelli2006limits,Creminelli2007estimators,Smith2011,Becker2012wmapRunningfNL}, and the formalism is mathematically equivalent to the ones described in Planck PNG papers \cite{PlanckCollaboration2013,PlanckCollaboration2015,PlanckCollaboration2018}. This way of presentation was chosen to clearly state the assumptions we make to write down the CMB bispectrum estimator and to draw parallels to the CMB power spectrum likelihoods. We consider only the CMB temperature here for simplicity, but the formalism can be extended to include polarization, as described in Appendix \ref{section:appendix_full_cmb_bispectrum_estimator}.

\subsection{CMB bispectrum likelihood}

CMB observations provide us with the spherical harmonic coefficients $a_{\ell m}$ of the temperature or polarization anisotropy maps. We are interested in their statistical properties, most of which are captured by their angular power spectrum estimated via
\begin{align}
    \hat{C}_\ell \equiv \sum_m \frac{1}{2\ell+1} a_{\ell m} a_{\ell m}^*.
\end{align}
This is a sufficient statistic only if the $a_{\ell m}$s are Gaussian distributed. Going one order beyond the power spectrum, we can construct an estimate for the bispectrum as
\begin{align}
    \hat{B}_\ells \equiv \sum_{m_j} \begin{pmatrix} \ell_1& \ell_2& \ell_3 \\ m_1& m_2& m_3 \end{pmatrix} \Bigl[  a_{\ell_1 m_1} a_{\ell_2 m_2} a_{\ell_3 m_3} - \left[ \langle a_{\ell_1 m_1} a_{\ell_2 m_2} \rangle_\mathrm{MC} \; a_{\ell_3 m_3} + (2~\mathrm{cyc.})  \right] \Bigr]. \label{eqn:observed_bispectrum}
\end{align}
The weights are the Wigner-3j symbols related to angular momentum conservation; $\hat{B}_\ells$ vanishes unless $\ell_1$, $\ell_2$ and $\ell_3$ form a triangle. The two terms in square brackets, cubic and linear in $a_{\ell m}$s, together estimate the full bispectrum ${\langle a_{\ell_1 m_1} a_{\ell_2 m_2} a_{\ell_3 m_3} \rangle}$. The linear term is necessary to reduce the variance of the estimate and keep it unbiased in the presence of an anisotropic sky mask, which introduces extra off-diagonal correlations in the power spectrum which correlate with $a_{\ell m}$s. Gaussian simulations are used for Monte-Carlo approximations of the full covariance appearing in the linear term. $\hat{B}_\ells$ can be understood as a summary statistic for the CMB 3-point functions under the assumption of statistical isotropy.

We define the theoretical bispectrum as the expected value of the estimate;
\begin{align}
    B^\mathrm{th}_\ells \equiv \langle \hat{B}_\ells \rangle 
    = h_\ells b^\mathrm{th}_\ells, \label{eqn:theory_bispectrum}
\end{align}
where $h_\ells$ is a geometric factor (\eqref{eqn:gaunt_in_terms_of_h} in Appendix \ref{section:appendix_identities}) and $b_\ells$ is the reduced bispectrum.

In order to write down the likelihood, we make the following approximations:
\begin{enumerate}
    \item $\{ \hat{B}_\ells \}_{\ell_1 \le \ell_2 \le \ell_3}$ are multivariate normal distributed.
    \item $\{ \hat{B}_\ells \}_{\ell_1 \le \ell_2 \le \ell_3}$ have a diagonal covariance matrix with diagonal entries  $(6 / \Delta_\ells) ~ C_{\ell_1} C_{\ell_2} C_{\ell_3}$.
\end{enumerate}
Here, the symmetry factor $\Delta_\ells = 6,3,1$ when there are 3, 2, 1 distinct values of $l_j$s, respectively.

The first approximation is justified by the central limit theorem, analogously to the power spectrum analysis. There are $O(\ell_1 \ell_2)$ independent terms appearing in \eqref{eqn:observed_bispectrum} for each $(\ell_1,\ell_2,\ell_3)$, so the weighted sum follows a near-Gaussian distribution. We note that this approximation can be inaccurate for a handful number of terms for which $\ell_1$ and $\ell_2$ are small (and so are $\ell_3$ by the triangle condition).

Omitting the non-diagonal terms in the covariance matrix is a choice to reduce computational complexity and was studied in \cite{Creminelli2006limits,Smith2011}. The values of diagonal entries then follow by Wick's theorem in the limit where $a_{\ell m}$s are only weakly non-Gaussian. The equivalent case for polarization is given in Appendix \ref{section:appendix_full_cmb_bispectrum_estimator}.

We note that, for a simple estimation of $\fNL$ using a single bispectrum template, the second approximation only affects the optimality of the estimator. The optimality of the estimator has been tested thoroughly in Planck analysis \cite{PlanckCollaboration2013,PlanckCollaboration2015,PlanckCollaboration2018} using multiple pipelines and is shown to be near-optimal. This approximation therefore would have little effect on our analysis. However, for future CMB surveys which will grant access to higher $\ell$s, this may no longer be the case. Contributions from the connected 4-point functions due to CMB lensing studied in \cite{Coulton2020bispectrumcovariance}, for example, are expected to be more significant, and the approximation above would underestimate the true covariance. Delensing has been proposed as a solution in \cite{Coulton2020bispectrumcovariance}.

Under the assumptions above, our statistical model is
\begin{align}
    \hat{\mathbf{B}} = \mathbf{B}^\mathrm{th} + \boldsymbol{\epsilon}, \label{eqn:bispectrum_base_model}
\end{align}
where the errors $\epsilon_\ells$ are multivariate normal with diagonal covariance. The CMB bispectrum likelihood is thus given by \footnote{We denote the likelihood as $\mathcal{L}(\boldsymbol{\theta} | \mathrm{data}) \equiv P(\mathrm{data} | \boldsymbol{\theta})$, which is the probability of observing the data given the model with parameters $\boldsymbol{\theta}$.}
\begin{align}
    \mathcal{L}(\mathbf{B}^\mathrm{th} | \hat{\mathbf{B}}) = A\; \mathrm{exp} \left[ -\frac{1}{2} \sum_{\ell_1 \le \ell_2 \le \ell_3} \frac{ \Delta_\ells }{ 6\; C_{\ell_1} C_{\ell_2} C_{\ell_3}} \left( \hat{B}_\ells - B^\mathrm{th}_\ells \right)^2 \right].  \label{eqn:bispectrum_likelihood_formal}
\end{align}
The normalisation constant $A = (2\pi)^{-n_{B}/2} \prod \left[ 6 C_{\ell_1} C_{\ell_2} C_{\ell_3}) / \Delta_\ells \right]^{-1/2}$, where $n_B$ denotes the number of ordered triplets $(\ell_1,\ell_2,\ell_3)$ in consideration.


\subsection{CMB bispectrum estimator of $\fNL$}

In power spectrum analyses, $\Lambda$CDM or other models provide theoretical predictions of $C_\ell (\boldsymbol{\theta})$s through Boltzmann solvers such as CAMB \cite{Lewis2000} and CLASS \cite{Blas2011class}, where $\boldsymbol{\theta}$ denotes the model parameters. The likelihood, combined with some priors on the parameters, then provides a posterior distribution which places bounds on the parameters. A similar analysis can be done for the bispectrum likelihood with any $n$-parameter model $\mathbf{B}^\mathrm{th}(\boldsymbol{\theta})$. However, since the signal-to-noise ratio of the CMB bispectrum is much smaller, the constraining power is often limited. Instead, we study some bispectrum templates that are well-motivated by models of the early universe and perform a simple linear fit, which constrains the amplitude of PNG. In our work, the cosmological parameters are fixed to their best-fit values at the power spectrum level, since their variation does not significantly affect the bispectrum amplitude \cite{PlanckCollaboration2015}.

The bispectrum of the curvature perturbations at the end of inflation is defined through 
\begin{align}
    \langle \zeta(\mathbf{k}_1) \zeta(\mathbf{k}_2) \zeta(\mathbf{k}_3) \rangle = (2\pi)^3 \delta^{(3)}(\mathbf{k}_1 + \mathbf{k}_2 + \mathbf{k}_3) B_\zeta (k_1, k_2, k_3).
\end{align}
Given one or more bispectrum templates $B_\zeta^{(i)}(k_1,k_2,k_3)$, we introduce some free parameters $\fNL^{(i)}$ to represent the amplitude of PNG;
\begin{align}
   B^\mathrm{th}_\ells (\mathbf{f}_\mathrm{NL}) &= \sum_{i} \fNL^{(i)} B^{(i)}_\ells = \sum_i \fNL^{(i)} h_\ells b^{(i)}_\ells,
\end{align}
where the reduced CMB bispectra are related to their primordial counterparts through
\begin{align}
    b^{(i)}_\ells &= \left( \frac{2}{\pi} \right)^3 \int dr dk_1 dk_2 dk_3 \; (r k_1 k_2 k_3)^2  B_\zeta^{(i)}(k_1,k_2,k_3) \prod_{j=1}^{3} \left[ j_{\ell_j}(k_j r) T_{\ell_j} (k_j) \right].
\end{align}
Here, $j_\ell(k)$ denotes the spherical Bessel function. The CMB transfer functions $T_\ell(k)$ are obtained from the background cosmology which is fixed as $\Lambda$CDM best-fit parameters to the CMB power spectrum likelihood. For Planck data, the bispectrum analyses are insensitive to this choice of background parameters \cite{PlanckCollaboration2013,PlanckCollaboration2015,PlanckCollaboration2018}. 

In most cases, a model of inflation predicts a bispectrum whose shape is approximated by a single template, so an independent analysis with a single $\fNL$ parameter suffices. However, if a class of models predicts bispectra that are expressed as linear combinations of two or more templates, then it is appropriate to have a joint analysis with multiple $\fNL$ parameters at once. Planck team provides joint constraints to the equilateral and orthogonal shapes \cite{PlanckCollaboration2018}, for example, since general single-field inflation models often yield a combination of the two.

The CMB bispectrum likelihood is then a function of $\fNL^{(i)}$s:
\begin{align}
    -2\ln \mathcal{L}(\mathbf{f}_\mathrm{NL} | \hat{\mathbf{B}}) &=  \mathrm{(const.)} - 2\sum_i S_i \fNL^{(i)} + \sum_{i,j} F_{ij} \fNL^{(i)} \fNL^{(j)}, \label{eqn:fNL_likelihood}
\end{align}
where we have defined
\begin{align}
    S_i &\equiv \sum_{\ell_1 \le \ell_2 \le \ell_3} \frac{ \Delta_\ells }{ 6\; C_{\ell_1} C_{\ell_2} C_{\ell_3}}  \hat{B}_\ells B^{(i)}_\ells \\
    &= \sum_{\ell_j, m_j} \frac{\mathcal{G}^\ells_{m_1 m_2 m_3} b^{(i)}_\ells }{6\; C_{\ell_1} C_{\ell_2} C_{\ell_3}} \Bigl[ a_{\ell_1 m_1} a_{\ell_2 m_2} a_{\ell_3 m_3} - \left[ \langle a_{\ell_1 m_1} a_{\ell_2 m_2} \rangle a_{\ell_3 m_3} + (2~\mathrm{cyc.})  \right] \Bigr],  \label{eqn:cmb_bispectrum_MLE_signal} \\
    F_{ij} &\equiv  \sum_{\ell_1 \le \ell_2 \le \ell_3} \frac{ \Delta_\ells }{ 6\; C_{\ell_1} C_{\ell_2} C_{\ell_3}}  B^{(i)}_\ells B^{(j)}_\ells \\
    &= \sum_{\ell_1, \ell_2, \ell_3} \frac{ h^2_\ells b^{(i)}_\ells b^{(j)}_\ells }{ 6\; C_{\ell_1} C_{\ell_2} C_{\ell_3}} .  \label{eqn:cmb_bispectrum_MLE_fisher}
\end{align}
We have replaced the sums over $\ell_1\le\ell_2\le\ell_3$ with the ones over $\ell_1,\ell_2,\ell_3$ using the symmetry factor $\Delta_\ells$ above. The Gaunt integral comes from the geometric factors given in \eqref{eqn:gaunt_in_terms_of_h}.

The CMB bispectrum estimator is the maximum likelihood estimator (MLE) of \eqref{eqn:fNL_likelihood}:
\begin{align}
    \widehat{\fNL}^{(i)} = \sum_j (F^{-1})_{ij} S_j . \label{eqn:cmb_bispectrum_estimator_MLE}
\end{align}
Assuming that the assumptions made above are valid, the MLE is unbiased so that $\langle \widehat{\fNL}^{(i)} \rangle = \fNL^{(i)}$. Furthermore, the estimator is optimal by the Gauss-Markov theorem; it has the smallest variance amongst all unbiased estimators constructed from $\hat{B}_\ells$. Its variance is then given by the Cramér-Rao bound which is expressed in terms of the Fisher information matrix $F_{ij}$ as
\begin{align}
    \mathrm{Cov}(\widehat{\mathbf{f}_\mathrm{NL}}) = F^{-1}.
\end{align}
Therefore, the marginal error on the parameter $\fNL^{(i)}$ is equal to $\sigma(\fNL^{(i)}) = (F^{-1})_{ii}$ (no summation implied). Note that this is in general different from an independent analysis with one bispectrum template, for which $\sigma(\fNL^{(i)}) = (F_{ii})^{-1}$ (no sum).

In practice, we use simulations (FFP10 end-to-end CMB maps \cite{PlanckCollaboration2015simulations,PlanckCollaboration2018hfi}) with Gaussian initial conditions to obtain the sample variance of $\fNL^{(i)}$, instead of directly using the Fisher error bar above. In a weakly non-Gaussian regime with a sufficient number of simulations, this variance accurately represents the variance of the estimator $\widehat{\fNL}$. The simulations are also used to approximate the full non-diagonal covariance $\langle a_{\ell_1 m_1} a_{\ell_2 m_2} \rangle_\mathrm{MC}$ appearing in \eqref{eqn:cmb_bispectrum_MLE_signal}.

\subsection{Beam, noise, sky mask, and lensing}

Our observations of the CMB are subject to having a finite beam width, discrete pixelization, instrumental noise, partial sky coverage, and CMB weak lensing. We outline here how these effects are taken into account.

The window function $W_\ell$ is a product of the beam and pixel window functions. The Planck CMB maps obtained through SMICA \cite{PlanckCollaboration2018component} have an effective beam full width at half maximum (FWHM) of $5$ arcmin for temperature and $10$ arcmin for polarization. Having a finite beam width limits the resolution and suppresses powers in high-$\ell$ multipoles. The beam window function is given by $\mathrm{exp}(-\ell(\ell+1)/2\sigma_\mathrm{beam}^2)$, where $\sigma_\mathrm{beam}=(\mathrm{FWHM})/\sqrt{8\ln 2}$, for temperature and polarization. Discrete pixelizations also has a similar effect of suppressing small-scale powers. Using the Healpix pixelization \cite{Gorski2005healpix} with $N_\mathrm{side}=2048$, the pixel window function is computed using the Python library \texttt{healpy} \cite{Zonca2019healpy}.

We obtain the power spectra of instrumental noise, $N_\ell$, from 300 end-to-end simulated noise maps \cite{PlanckCollaboration2015simulations,PlanckCollaboration2018hfi}. This yields consistent results when compared with the SMICA post-component-separation noise. We further assume that the temperature and E-mode polarization noises are uncorrelated and that their three-point functions vanish. The noise terms therefore only appear in $C_\ell^\mathrm{TT}$ and $C_\ell^\mathrm{EE}$.

The sky masks for temperature and polarization maps with $f_\mathrm{sky}^\mathrm{T}=0.779$ and $f_\mathrm{sky}^\mathrm{E}=0.781$ \cite{PlanckCollaboration2018component} have been used for our analysis. Assuming statistical isotropy of the Universe, the expectation value $\langle \cdot \rangle$ gains a factor of $f_\mathrm{sky}$. Furthermore, sharp discontinuities in the CMB map due to sky masks have been shown to cause numerical issues for the bispectrum analysis \cite{Gruetjen2017inpainting}. In particular, the small-scale powers boosted by sharp cut-sky correlate with the large-scale multipoles from the mask shape, inducing a significant bias of local-type bispectrum. We therefore adopt the method of \cite{Gruetjen2017inpainting} and inpaint the CMB maps around the edges of sky masks via linear isotropic diffusion. Inpainting smooths the mask around the edges and hence transfers the mask-induced small-scale CMB powers to larger scales.

The CMB gets weakly lensed by matter as it travels from the last scattering surface to us. This weak lensing affects the CMB power spectrum by smoothing out the acoustic peaks slightly \cite{Dodelson2003textbook}. For the CMB bispectrum, small-scale powers induced by lensing correlate with the integrated Sachs-Wolfe (ISW) contributions to the large-scale modes, which creates a bias in the squeezed shape \cite{Hu2000cmblensing,Hanson2009lensingISW}. The lensing-ISW bias is given by \cite{Lewis2011lensing}
\begin{align}
    b^\mathrm{lensing-ISW}_\ells = \frac{1}{2} \left[ \ell_1(\ell_1 + 1) - \ell_2(\ell_2 + 1) + \ell_3 (\ell_3 + 1) \right] \tilde{C}^\mathrm{TT}_{\ell_1} C^{\mathrm{T}\phi}_{\ell_3} + (5\;\mathrm{perms.}),
\end{align}
where $C^{\mathrm{T}\phi}_\ell$ is the temperature and lensing potential cross power spectra. A tilde above $C_\ell$ signifies that it is a lensed quantity. We explicitly include the bias above in our analysis.

In summary, the statistical model \eqref{eqn:bispectrum_base_model} is modified as the following:
\begin{align}
    \hat{B}_\ells = f_\mathrm{sky} W_{\ell_1} W_{\ell_2} W_{\ell_3} \left( B^\mathrm{th}_\ells + B^\mathrm{lensing-ISW}_\ells \right) + \epsilon_\ells,
\end{align}
where the error has a vanishing mean and a diagonal covariance equal to
\begin{align}
    \mathrm{Cov}(\epsilon_\ells) = f_\mathrm{sky} \; \mathrm{diag} \left( \frac{6}{\Delta_\ells} (W^2_{\ell_1} C_{\ell_1} + N_{\ell_1}) (W^2_{\ell_2} C_{\ell_2} + N_{\ell_2}) (W^2_{\ell_3} C_{\ell_3} + N_{\ell_3}) \right),
\end{align}
and $B^\mathrm{lensing-ISW}_\ells = h_\ells b^\mathrm{lensing-ISW}_\ells$.

\section{High-resolution CMB bispectrum estimator} \label{section:high-resolution_CMB_bispectrum_estimator}

Computation of the CMB bispectrum estimator \eqref{eqn:cmb_bispectrum_estimator_MLE} can be prohibitively expensive; the most naive method would require summing over $O(\ell_\mathrm{max}^5) \sim O(10^{16})$ terms. All existing techniques rely on one or more tricks and/or assumptions to simplify the computation process. The KSW estimator \cite{Komatsu2005} utilizes separable bispectrum templates for which the three-dimensional integral $dk_1 dk_2 dk_3$ splits into a product of three separate one-dimensional integrals. The formalism is fast and efficient but is restricted to a limited range of separable templates. The Modal estimator \cite{Fergusson2010general,Fergusson2012,Fergusson2014} expands the primordial and late-time bispectra in terms of separable basis functions. The bispectrum information is compressed with respect to the basis and stored. This allows fast and thorough analyses of general bispectrum templates. Lastly, the binned bispectrum estimator \cite{Bucher2010,Bucher2016} bins the bispectrum into different $\ell$ bins, which makes the total size more computationally tractable with minimal loss of optimality. We refer to \cite{PlanckCollaboration2015} and references therein for detailed reviews on the CMB bispectrum estimation.

In this section, we introduce our novel, independent CMB bispectrum estimator \best, which combines ideas from the KSW estimator \cite{Komatsu2005} and the Modal estimator \cite{Fergusson2010general,Fergusson2012}. While being computationally more expensive than the two, \best\ combines the best of both worlds to be general and efficient, and has the flexibility in the choice of basis for high-resolution analyses.

\subsection{\best\ formalism}

In \best, the primordial bispectra are expanded using separable basis functions similar to Modal \cite{Fergusson2010general}, followed by the compression of the CMB bispectrum information with respect to this basis in a way similar to the KSW estimator \cite{Komatsu2005}. We detail the formalism here.

Given a choice of one-dimensional mode functions $q_p(k)$, a given primordial bispectrum template is expanded as
\begin{align}
	(k_1 k_2 k_3)^2 B_\zeta (k_1, k_2, k_3) =  \sum_{p_1, p_2, p_3} \alpha_{p_1 p_2 p_3} \; q_{p_1}(k_1) q_{p_2}(k_2) q_{p_3}(k_3). \label{eqn:BEST_basis_expansion}
\end{align}
The expansion above holds for bispectrum templates that are accurately represented using the basis functions $Q_{p_1 p_2 p_3}(k_1,k_2,k_3) \equiv q_{p_1}(k_1) q_{p_2}(k_2) q_{p_3}(k_3)$. \footnote{The truncation error is small as long as $B$ is within or close to the function space spanned by the basis functions $\{ q_{p_1}(k_1) q_{p_2}(k_2) q_{p_3}(k_3) \}$. } Note that, by symmetry, the above is equivalent to having the sum over $p_1\ge p_2\ge p_3$ with an additional symmetry factor $\Delta_{p_1 p_2 p_3}$. This convention will be used in the next section, where different choices of mode functions and the expansion procedure are discussed. Here, there are no restrictions to each $p_j$ which runs from $1$ to $p_\mathrm{max}$, the number of mode functions.

The basis functions are products of terms that depend on only one of the $k$s. The reduced bispectrum simplifies thanks to this separability:
\begin{align}
	b_\ells = \sum_{p_j} \alpha_{p_1 p_2 p_3} \int dr \; \tilde{q}_{p_1}(\ell_1, r) \tilde{q}_{p_2}(\ell_2, r) \tilde{q}_{p_3}(\ell_3, r), \label{eqn:reduced_bispectra_projected_modes}
\end{align}
where the projected mode functions are defined as
\begin{align}
	\tilde{q}_{p}(\ell, r) \equiv \frac{2r^\frac{2}{3}}{\pi} \int dk \; q_p(k) T_\ell (k) j_\ell (kr).  \label{def:projected_modes}
\end{align}
In the Modal estimator \cite{Fergusson2010general}, another set of mode functions are introduced in the $\ell$ space to form a late-time basis so that $b_\ells = \sum_{p_j} \tilde{\alpha}_{p_1 p_2 p_3} \tilde{Q}_{p_1 p_2 p_3}(\ells) $. This effectively removes the line-of-sight integral $\int dr$ appearing in \eqref{eqn:reduced_bispectra_projected_modes} and reduces the computational complexity by a couple of orders of magnitude. The expansion accurately approximates most bispectrum templates; the correlation levels between the template and the basis expansion vary between $\sim0.95$ and $\sim0.99$ for Modal analysis in Planck \cite{PlanckCollaboration2013}. However, for bispectrum templates motivated by feature or resonance models, the Modal estimator had limited coverage compared to the KSW-type estimators \cite{PlanckCollaboration2018}. In \best, we do not perform this second late-time basis expansion step and instead follow a KSW-like formalism to compute the estimator exactly, albeit with increased computational complexity.

We define the filtered maps from $a_{\ell m}$s as
\begin{align}
	M^{(i)}_p (\hat{\vv{n}}, r) \equiv \sum_{\ell,m} \frac{\tilde{q}_p (l,r)}{C_l} a_{\ell m} Y_{\ell m} (\hat{\vv{n}}). \label{def:filtered_maps}
\end{align}
The observed map corresponds to $i=0$ in our notation above, while $i=1,\cdots,N_\mathrm{sim}$ signify the FFP10 end-to-end CMB and noise map numbers \cite{PlanckCollaboration2015simulations,PlanckCollaboration2018hfi} under Gaussian initial conditions. As described in Section \ref{section:CMB_bispectrum_estimation}, these simulations are used for two purposes: first to approximate the full covariance matrix $\langle a_{\ell_1 m_1} a_{\ell_2 m_2} \rangle_\mathrm{MC}$ appearing in the linear term of \eqref{eqn:cmb_bispectrum_MLE_signal}, and second to form a set of $\fNL$ estimates which is used as a null test to evaluate the statistical significance of $\fNL$ estimated from observations.

The CMB bispectrum estimator for a single bispectrum template can then be written as
\begin{align}
	\widehat{\fNL}^{(i)} &= \frac{1}{F} \sum_{\ell_j,m_j} \frac{\mathcal{G}^\ells_{m_1 m_2 m_3} b_\ells}{ 6C_{\ell_1} C_{\ell_2} C_{\ell_3}} \left[ a_{\ell_1 m_1}^{(i)} a_{\ell_2 m_2}^{(i)} a_{\ell_3 m_3}^{(i)} -  \left( \frac{1}{N_\mathrm{sim}-1} \sum_{j \neq i} a^{(j)}_{\ell_1 m_1} a^{(j)}_{\ell_2 m_2}  a^{(i)}_{\ell_3 m_3} + (\mathrm{2\ perms.}) \right)  \right].		\label{eqn:bispectrum_estimator_standard} \\
        &= \frac{1}{6F} \sum_{p_1,p_2,p_3} \alpha_{p_1 p_2 p_3} \left[ \beta^{\mathrm{cub},(i)}_{p_1 p_2 p_3} - 3 \beta^{\mathrm{lin},(i)}_{p_1 p_2 p_3} \right], \label{eqn:bispectrum_estimator_alpha_beta}
\end{align}
where
\begin{align}
	\beta^{\mathrm{cub},(i)}_{p_1 p_2 p_3} &\equiv \int dr \int d^2\hat{\vv{n}} \; M^{(i)}_{p_1} (\hat{\vv{n}},r) M^{(i)}_{p_2} (\hat{\vv{n}},r) M^{(i)}_{p_3} (\hat{\vv{n}},r),	\label{def:beta_cubic} \\
	\beta^{\mathrm{lin},(i)}_{p_1 p_2 p_3} &\equiv \frac{1}{N_\mathrm{sim}-1} \sum_{j \neq i} \int dr \int d^2\hat{\hat{\vv{n}}} \; M^{(j)}_{p_1} (\hat{\vv{n}},r) M^{(j)}_{p_2} (\hat{\vv{n}},r) M^{(i)}_{p_3} (\hat{\vv{n}},r). \label{def:beta_linear}
\end{align}
Computing the $\beta^\mathrm{cub}$ and $\beta^\mathrm{lin}$ are the most time-consuming step of \best. Once they are computed, any given bispectrum template can be constrained instantly after a primordial basis expansion, which normally takes less than a minute on a laptop.

The normalisation, directly related to the Fisher information, is obtained similarly by exploiting separability and the relation \eqref{eqn:h2_in_terms_of_legendre_polys} for $h^2_\ells$;
\begin{align}
	F = \frac{1}{6} \sum_{p_j, p'_j} \alpha_{p_1 p_2 p_3} \Gamma_{p_1 p_2 p_3, p'_1 p'_2 p'_3} \alpha_{p'_1 p'_2 p'_3}, \label{eqn:normlisation_from_gamma}
\end{align}
where
\begin{align}
	\Gamma_{p_1 p_2 p_3, p'_1 p'_2 p'_3} &\equiv \int dr \int dr' \int d\mu \; \gamma_{p_1 p'_1}(\mu, r, r') \gamma_{p_3 p'_3}(\mu, r, r') \gamma_{p_3 p'_3}(\mu, r, r'), 	\label{def:gamma_dr_integrals}\\
	\gamma_{p p'}(\mu, r, r') &\equiv \sum_\ell \frac{2\ell+1}{(8\pi)^{1/3} C_\ell} \tilde{q}_{p}(\ell,r) \tilde{q}_{p'}(\ell,r') P_\ell(\mu).
\end{align}
Here, $P_\ell(\mu)$'s denote the Legendre polynomials.

Lastly, we summarise the key differences of \best\ with two of the main methods used in Planck in Table \ref{table:comparison_with_ksw_modal}. Note that the binned bispectrum estimator \cite{Bucher2010} is omitted in the table due to its dissimilarity with \best, but it is also one of the main approaches used extensively in Planck analyses.

\begin{table}[htbp!]
    \caption{Comparison of \best\ with the two conventional bispectrum estimators \cite{Komatsu2005,Fergusson2010general} used in Planck analyses \cite{PlanckCollaboration2013,PlanckCollaboration2015,PlanckCollaboration2018}. The core ideas for data/information reduction for different approaches are detailed in the text. $B_\zeta(k_1,k_2,k_3)$ and $B^\mathrm{th}_\ells$ correspond to the primordial bispectrum template and its late-time harmonic counterpart, respectively.  }
	\centering
	\label{table:comparison_with_ksw_modal}
	\renewcommand{\arraystretch}{2.0} 
	\begin{tabular}{P{20mm}P{40mm}P{30mm}P{30mm}P{30mm}}
            \toprule
            & \multirow{2}{*}{\centering Core idea} & \multicolumn{2}{c}{Estimation accuracy} & \multirow{2}{*}{\parbox{30mm}{\centering Relative computational complexity \\ (rough estimate) }} \\
            \cmidrule{3-4}
            & & Separable templates & Non-separable templates & \\
            \midrule
            KSW \lbrack Komatsu \textit{et al.} 2005 \rbrack & Use separable templates & Exact & Not applicable & $\sim 1$ per model \\
            \midrule
            Modal \lbrack Fergusson \textit{et al.} 2010\rbrack & Expand $B_\zeta(k_1,k_2,k_3)$ and $B_\ells^\mathrm{th}$ using separable basis functions & As good as the $B_\ells^\mathrm{th}$ expansion & As good as the $B_\ells^\mathrm{th}$ expansion & $\sim 30$ \\
            \midrule
            \best \quad \lbrack this work\rbrack & Expand $B_\zeta(k_1,k_2,k_3)$ using separable basis functions & Exact & As good as the $B_\zeta(k_1,k_2,k_3)$ expansion & $\sim 10,000$ \\
            \bottomrule
        \end{tabular}
\end{table}

\subsection{Basis expansions} \label{section:basis_expansions}

Computing the quantities $\beta$ \eqref{def:beta_cubic} and $\Gamma$ \eqref{def:gamma_dr_integrals} are computationally expensive but need to be performed only once per data and basis set. Afterwards, \best\ can rapidly constrain bispectrum shapes of interest through a two-step procedure: (1) expand the shape function with respect to a separable basis, and (2) compute $\fNL$ and $\sigma(\fNL)$ via some simple matrix multiplications of the expansion coefficients with $\beta$ and $\Gamma$. We provide the precomputed data for various basis sets together with our public code (detailed in Section \ref{section:public_release}) for the two steps.

We expand a given bispectrum shape $S(k_1,k_2,k_3) = (k_1 k_2 k_3)^2 B_\zeta (k_1,k_2,k_3)$ in terms of the basis functions as follows. First, we simplify our basis functions by symmetrising over $(k_1,k_2,k_3)$;
\begin{align}
    \sum_{p_1, p_2, p_3} \alpha_{p_1 p_2 p_3} q_{p_1} (k_1) q_{p_2} (k_2) q_{p_3} (k_3) &= \sum_{p_1, p_2, p_3} \alpha_{p_1 p_2 p_3} \; \frac{1}{6} \Bigl[ q_{p_1} (k_1) q_{p_2} (k_2) q_{p_3} (k_3) + (\mathrm{5\ perms.})\Bigr] \\
    &= \sum_{n\leftrightarrow (p_1,p_2,p_3)} \alpha_n Q_n (k_1, k_2, k_3),
\end{align}
where $n$ is an index mapped one-to-one with a triplet $(p_1,p_2,p_3)$ satisfying $p_1\ge p_2 \ge p_3$. Note that ${\alpha_n = \Delta_{p_1 p_2 p_3} \alpha_{p_1 p_2 p_3}}$, which contains an extra symmetry factor.

To obtain $\alpha_n$, we solve the following linear equation:
\begin{align}
    \sum_{n'} \left< Q_n,\; Q_{n'} \right>  \alpha_{n'} &= \left< Q_n,\; S \right>, \quad \mathrm{where}    \label{eqn:basis_expansion_QQ_QS} \\
    \left<f,\; g\right> &\equiv \int_{\tetra} d^3\vv{k}\; w(\vv{k}) f(\vv{k}) g(\vv{k}).  \label{eqn:tetrapyd_inner_product}
\end{align}
The inner product $\langle\cdot,\cdot\rangle$ of two functions over the tetrapyd domain is defined for some weight function $w(\vv{k})$. Mathematically, solving \eqref{eqn:basis_expansion_QQ_QS} for $\alpha$ is equivalent to finding an orthogonal projection of $S$ into the function space spanned by the basis functions $\{ Q_n \}$. The truncation error of this basis representation comes from the component of $S$ that is perpendicular to this function space.

For small basis sizes, we can directly invert the matrix $\Gamma_{n n'} \equiv \langle Q_n,\; Q_{n'} \rangle$ to solve the linear equation for $\alpha$. However, this can become numerically unstable for larger bases since some basis functions become more degenerate, which degrades the condition number of $\Gamma$.\footnote{Note that even though the Legendre mode functions are orthogonal in one dimension, the three-dimensional basis functions constructed are no longer orthogonal on the (non-separable) tetrapyd domain.}  Instead of a direct inversion, we use the conjugate gradient method \cite{Hestenes1952conjugategradient} to obtain an approximate solution for \eqref{eqn:basis_expansion_QQ_QS} with only marginal residual errors. This iterative method is applicable here because $\Gamma$ is symmetric and positive definite by construction, and it can retrieve the exact solution when the matrix is accurately invertible. We found that the iterative algorithm implemented in the Python library \texttt{scipy} \cite{SciPy} is numerically stable and efficient for our purposes.

\subsection{Basis function choices} \label{section:basis_function_choices}

One of the greatest strengths of \best\ lies in its flexibility with the choice of mode functions from which the basis set is constructed. Adopting a small set of specific mode functions provides fast and precise results for some specific bispectrum templates of interest. On the other hand, a general basis set allows us to constrain a broad range of inflationary models simultaneously but requires more computational resources upfront. In this section, we describe three types of basis sets studied in this work.

The first and simplest basis set consists of monomial mode functions of the form
\begin{align}
	q_p(k) = k^{p-1}, \quad \mathrm{for} \quad p = 0, 1, 2, 3.     \label{eqn:monomials_mode_functions}
\end{align}
The three most standard bispectrum templates\textemdash local, equilateral, and orthogonal\textemdash can be expressed as a sum of separable terms above. For example, the local template, for example, is given by
\begin{align}
        S^\mathrm{local}(k_1 k_2 k_3) &\equiv (k_1 k_2 k_3)^2 B^\mathrm{local}_\Phi (k_1, k_2, k_3) = 2 A^2 \left[ \frac{k_1^2}{k_2 k_3} + \frac{k_2^2}{k_3 k_1} + \frac{k_3^2}{k_1 k_2}  \right].  \label{eqn:local_shape_function}
\end{align}
where $A$ is the primordial gravitational potential ($\Phi$) power spectrum amplitude, which relates to the usual scalar (curvature $\zeta$) power spectrum amplitude through $A = 2\pi^2 (3/5)^2 A_\mathrm{s}$ due to different conventions \footnote{The factor of $(3/5)^2$ comes from the fact that $\Phi = \frac{3}{5} \zeta$ at superhorizon scales. The other factor of $2\pi^2$ is from the relation between the power spectrum $P(k)$ and the dimensionless power spectrum $\mathcal{P}(k)$: $P(k) = (2\pi^2/k^3) \mathcal{P}(k)$. }. Decomposition coefficients $\alpha_{p_1 p_2 p_3}$ for the local template are given exactly and have three non-zero components: $\alpha_{300} = \alpha_{030} = \alpha_{003} = 2A^2$. Coefficients for the equilateral and orthogonal templates are obtained in a similar fashion. \footnote{In practice, we account for a non-unit $n_\mathrm{s}$ by modifying the basis as $q_p(k) = k^2 \left[ k_* \left( k/k_* \right)^{(4-n_\text{s})/3} \right]^{p-3}$ for $p=0,1,2,3$. The pivot scale $k_* = 0.05\mathrm{Mpc}^{-1}$ so that $\mathcal{P}_\zeta (k) = A_\mathrm{s} k^{n_\mathrm{s}-1}$.}

Choosing the monomial functions and using the exact $\alpha$s above render \best\ to be completely equivalent to the KSW estimator. We will refer to this basis set as `Monomials'.

\begin{figure*}
	\centering    
	\includegraphics[width=1.0\textwidth]{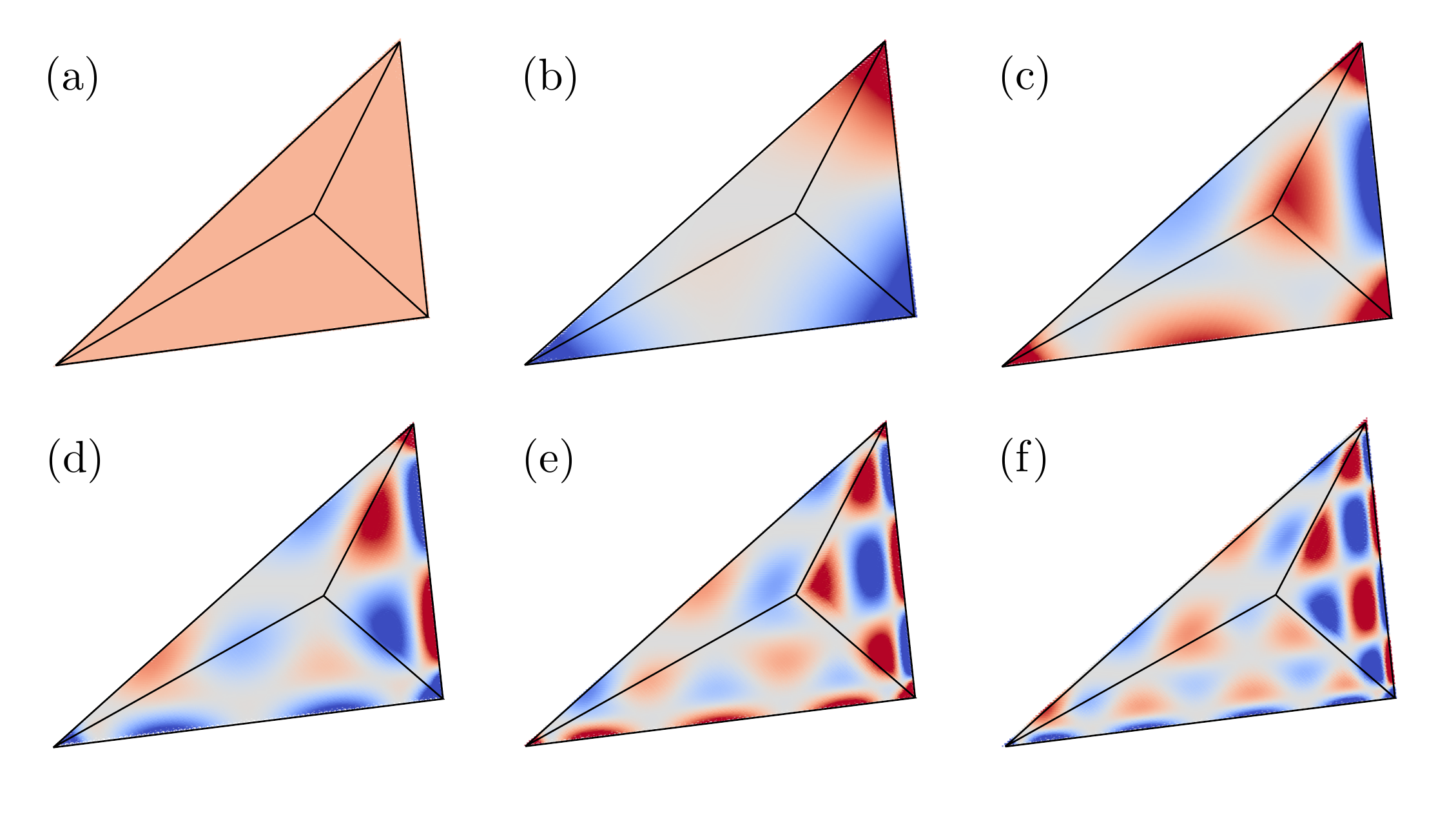}
	\caption{Some examples of our Legendre basis functions, evaluated on a sliced tetrapyd domain with $k_1\ge k_2\ge k_3$. The basis functions are defined as $Q_{p_1 p_2 p_3}(k_1,k_2,k_3) \equiv q_{p_1}(k_1) q_{p_2}(k_2) q_{p_3}(k_3)$, where $q_p(k)$ are defined in \eqref{eqn:legendre_mode_functions}. Here we plot $p_1 = p_2 = p_3 = p$, where $p$ equals (a) $1$, (b) $2$, (c) $3$, (d) $4$, (e) $5$, and (f) $6$. A single colour map is used across the plots: red and blue correspond to $+1$ and $-1$, respectively. Note that this figure has been borrowed from the thesis of \cite{Sohn2022thesis}. }	
	\label{fig:Legendre_basis_functions_3D}
\end{figure*}

For general analyses on a wide range of bispectrum templates, we restrict the $k$ range to $[\kmin,\kmax]$ and define the following mode functions;
\begin{align}
    q_0(k) &= k^{n_\mathrm{s} - 2}, \\
    q_p(k) &= P_p (\mu(k)), \quad p=1,2,\cdots,p_\mathrm{max}-1, \quad \mathrm{where}\quad \mu(k) \equiv -1 + \frac{2 (k - \kmin)}{\kmax-\kmin} . \label{eqn:legendre_mode_functions}
\end{align}
Many bispectrum templates have terms that depend on $k^{-1}$ which is captured by $q_0(k)$. The number of modes, $p_\mathrm{max}$, determines the maximum order of the Legendre polynomials $P_p(\mu)$ included. Our baseline analysis uses $p_\mathrm{max}=10$, while the high-resolution one has $p_\mathrm{max}=30$. The latter yields $4960$ basis functions in total after accounting for symmetries. We refer to this basis set as `Legendre'. Some of the basis functions in the Legendre basis set are shown in Figure \ref{fig:Legendre_basis_functions_3D}.

As promised in the name `high-resolution CMB bispectrum estimator', \best\ allows targeted analysis on complex, non-separable bispectrum shapes with high-frequency oscillations, which has not been constrained so far due to computational challenges. Given an oscillation frequency $\omega_*$ of interest, a targeted oscillatory basis can be constructed as the following:
\begin{align}
    q_0(k) &= k^{n_\mathrm{s} - 2} \sin(\omega_* k), &&\quad q_1(k) = k^{n_\mathrm{s} - 2} \cos(\omega_* k), && \\
    q_{2p}(k) &= P_p (\mu(k)) \sin(\omega_* k), &&\quad q_{2p+1}(k) = P_p (\mu(k)) \cos(\omega_* k), && \quad p=1,2,\cdots,(p_\mathrm{max}/2)-1. \label{eqn:sine_legendre_mode_functions}
\end{align}
This basis set can be thought of as a tensor product between the Legendre basis set and $\{\sin(\omega_* k), \cos(\omega_* k) \}$. A bispectrum shape with linearly-spaced oscillations and some envelope function $f$ can be rewritten as
\begin{align}
    S(k_1, k_2, k_3) = f(k_1,k_2,k_3) \sin(\omega_* ( k_1 + k_2 + k_3) + \phi) = \mathrm{Im}\left[ e^{i\phi} f(k_1,k_2,k_3) e^{i\omega_* k_1} e^{i\omega_* k_2} e^{i\omega_* k_3} \right].
\end{align}
Therefore, the expansion coefficients of $S$ with respect to the targeted oscillatory basis can be obtained by first expanding the envelope $A$ using a Legendre basis of order $p_\mathrm{max}/2$ and then taking a tensor product with the oscillatory part: $-\alpha_{000}=\alpha_{011}=\alpha_{101}=\alpha_{110}=\cos\phi$ and $-\alpha_{001}=-\alpha_{010}=-\alpha_{100}=\alpha_{111}=\sin\phi$.

\section{Numerical integration over tetrapyd} \label{section:numerical_integration_over_tetrapyd}

In this section, we present our novel method for a precise and efficient numerical integration over the `tetrapyd' volume, which appears frequently in bispectrum analyses including \best. The method shows excellent performance, especially for integrands that are well approximated by polynomials, showing orders of magnitude improvement in precision with much fewer evaluation points compared to simple trapezoidl rule, as will be discussed below. Our Python code \tetraquad\; is publicly available and can be found in \texttt{https://github.com/Wuhyun/Tetraquad}.

\subsection{Tetrapyd quadrature}

The primordial bispectrum is defined on a `tetrapyd' domain specified by triangle inequalities and the observational limits of the $k$ range;
\begin{align}
    \tetra(\kmin,\kmax) \equiv \{(k_1,k_2,k_3) : 2\max \{k_1, k_2, k_3\} \le k_1 + k_2 + k_3 \;\mathrm{and}\; \kmin \le k_1, k_2, k_3 \le \kmax \}. 
\end{align}
Volume integrals over this domain are not separable; they cannot be rewritten as three independent one-dimensional integrals, unlike the integral over the cube $[\kmin,\kmax]^3$. It is therefore difficult to simplify these integrals without introducing one or more extra integration variables.

The simplest way to numerically compute the volume integrals over the tetrapyd domain is, therefore, to approximate it with a weighted sum of the integrand evaluated at a finite number of points:
\begin{align}
    \int_{\tetra(\kmin,\kmax)} dV \; f(\vv{k}) \approx \sum_{n=1}^{N} w_n f(\vv{x}_n).  \label{eqn:quadrature_definition}
\end{align}
Such a method of numerical integration is often referred to as quadrature, or cubature in this case since it is a volume integral \cite{Ueberhuber1997numericalComputation}.

One of the simplest quadratures in one dimension is the trapezoidal rule with a uniformly spaced grid. Similarly, one can create a uniform grid inside the three-dimensional tetrapyd. Each grid point is weighted proportional to what fraction of its voxel (volume pixel: a cube with the grid point in the centre and shares sides with neighbouring grid points) intersects with the tetrapyd. This approach can yield robust results for various integrands and is sensitive to sharp local oscillations if there are any. However, it is computationally expensive to achieve high numerical precision using this method, because the number of grid points required for a three-dimensional volume scales up rapidly with the grid density. 

Efficient quadrature rules for various three-dimensional volumes including spheres, tetrahedra and pyramids have been studied (e.g., \cite{Ueberhuber1997numericalComputation,jaskowiec2021pyramidCubature}), but not for a rather complex shape of tetrapyd, to the authors' knowledge. Inspired by the Gaussian quadrature rules that yield exact results for the first $M$ orthogonal polynomials, we seek an (approximate) tetrapyd quadrature satisfying the following conditions:
\begin{enumerate}
    \item Polynomials $P(k_1,k_2,k_3) = k_1^{p} k_2^{q} k_3^{r}$ are evaluated almost exactly for $p+q+r < M$, for some $M$
    \item The nodes $\{\vv{x}_n\}$ and weights $\{w_n\}$ are invariant under the permutations of $(k_1, k_2, k_3)$, which are the symmetries that the tetrapyd volume enjoys
    \item The weights are non-negative ($w_n \ge 0$) for numerical stability 
\end{enumerate}

Note that a quadrature rule over $\tetra(\kmin,\kmax)$ can be easily obtained from that of $\tetra(\kmin/\kmax,1)$ after a suitable rescaling. We define $\kratio \equiv \kmin / \kmax$. Next, by symmetry,
\begin{align}
    \int_{\tetra(\kratio,1)} dV \; f(k_1,k_2,k_3) &= \frac{1}{6} \int_{\tetra(\kratio,1)} dV \; \left[ f(k_1,k_2,k_3) + 5\; \mathrm{perms.} \right] \\
    &= \int_{\tetrasix(\kratio,1)} dV \; \left[ f(k_1,k_2,k_3) + 5\; \mathrm{perms.} \right],
\end{align}
where
\begin{align}
    \tetrasix(\kratio,1) \equiv \{(k_1,k_2,k_3) : k_1 \le k_2 + k_3 \;\mathrm{and}\; \kratio \le k_3 \le k_2 \le k_1 \le 1 \}.
\end{align}
It is therefore sufficient to find a quadrature rule for the symmetric functions $f(k_1,k_2,k_3)$ over $\tetrasix(\kratio,1)$. The quadrature rule can be extended symmetrically anytime to cover $\tetra(\kratio,1)$ and satisfy the second condition above.

\subsection{Orthogonal polynomials}

To find a quadrature rule that exactly evaluates the integrals of polynomials over a tetrapyd, we first need an analytic formula for the integrals. We make use of the following expression for the integrals of $k_1^{p} k_2^{q} k_3^{r}$ over a unit tetrapyd:
\begin{align}
    \int_{\tetra(0,1)} dV \; k_1^{p} k_2^{q} k_3^{r} = \frac{1}{(p+1)(q+1)(r+1)} - \left[ \frac{\Gamma(1+q) \Gamma(1+r)}{(3+p+q+r)\Gamma(3+q+r)} + (2\; \mathrm{perms.}) \right], \label{eqn:tetra_analytic_expression}
\end{align}
where $\Gamma(n)$ denotes the gamma function. We have also derived (by hand) the full analytical expression for $\tetra(\kratio,1)$, which generalises \eqref{eqn:tetra_analytic_expression}. This result can be found in Appendix \ref{section:appendix_analytic_tetrapyd_formula}.

Next, we define an inner product over the tetrapyd as
\begin{align}
    \langle f, g \rangle_{\tetra(\kratio,1)} \equiv \int_{\tetra(\kratio,1)} dV \; f(\vv{k}) g(\vv{k}).
\end{align}
It is possible to use different weight functions for the integral, but the analytic formulae provided here will work only if they are of form $(k_1 k_2 k_3)^a$. We orthogonalise and normalise the symmetric polynomials with respect to this inner product using the modified Gram-Schmidt process (MGS). In our public code \tetraquad, MGS has been modified further to be slower but numerically more stable. The first four orthonormalised symmetric polynomials over $\tetra(0.1,1)$, for example, are given as follows:
\begin{align}
    P_0 (k_1, k_2, k_3) &= 1.4540, \\
    P_1 (k_1, k_2, k_3) &= 9.4663 \;\frac{k_1+k_2+k_3}{3} - 5.6334, \\
    P_2 (k_1, k_2, k_3) &= 43.945 \;\frac{k_1^2+k_2^2+k_3^2}{3} - 51.129 \;\frac{k_1+k_2+k_3}{3} + 12.439, \\
    P_3 (k_1, k_2, k_3) &= 40.363 \;\frac{k_1 k_2 + k_2 k_3 + k_3 k_1}{3} -13.116 \;\frac{k_1^2+k_2^2+k_3^2}{3} - 29.586 \;\frac{k_1+k_2+k_3}{3} + 8.3660. 
\end{align}
By orthogonality, all $P_d$ for $d>0$ integrate to zero over the tetrapyd. Therefore, an exact quadrature rule of order $M$ should satisfy
\begin{align}
    \sum_n w_n P_d(\vv{x}_n) = 0 \; \mathrm{for} \; 0 < d < M \quad \mathrm{and} \quad \sum_n w_n = \mathrm{vol}(\tetra(\kratio,1))
\end{align}
Note that the above would guarantee all symmetric polynomials of total order less than that of $P_M$ to be evaluated exactly.

\subsection{Finding approximate quadrature}

In this work, we find an approximate quadrature rule by fixing some grid points $\{ \vv{x}_n \}$ and solving the following non-negative least squares problem (NNLS):
\begin{align}
    \mathrm{Minimise} \quad L(\vv{w}) = \| P \vv{w} - \vv{a} \|^2 \quad \mathrm{subject \; to} \quad w_n \ge 0,
\end{align}
where the $M\times N$ matrix $P$ satisfies $P_{dn} = P_d(\vv{x}_n)$, $\vv{w}=(w_1,w_2,\cdots,w_N)^\mathrm{T}$, and $\vv{a}=(\mathrm{vol}(\tetra(\kratio,1)),0,0,\cdots,0)^\mathrm{T}$. Having $L(\vv{w})=0$ would mean that the polynomials $P_0, \cdots, P_{M-1}$ are evaluated exactly using the quadrature. In our code, we use the Python library \texttt{scipy} \cite{SciPy} to solve NNLS using an active set method \cite{lawson1995lstsq}. We will refer to this novel quadrature rule on tetrapyd as `Tetraquad'.

The parameters $M$ and $N$ are free for us to choose. The numerical precision of the approximate quadrature mainly depends on $M$, as it dictates the maximum order of polynomials the quadrature can handle with guaranteed accuracy. On the other hand, $N$, the number of grid points (and weights), should be sufficiently large so that there are enough free variables $w_n$ to minimise the error $L(\vv{w})$.

We have chosen to solve the optimisation problem above instead of inverting directly ($\vv{w} = P^{-1} \vv{a}$) for two reasons. First, the matrix $P$ is often singular and therefore challenging to invert. We found that this is especially problematic for grid points that are regularly spaced inside the tetrapyd volume. Second, the direct inversion does not guarantee our requirement that the weights are non-negative, which is crucial for the numerical stability.

Solving the non-negative least squares problem has another advantage; the optimal solution often leaves many of the $w_n$'s exactly equal to zero. In fact, we found that less than 10\% of the weights remained non-zero in most cases when using a grid uniformly spaced within the tetrapyd. This allows us to drop the grid points that do not contribute to the quadrature, which in turn effectively decreases $N$ without degrading accuracy.

\begin{figure}
    \includegraphics[width=0.4\textwidth]{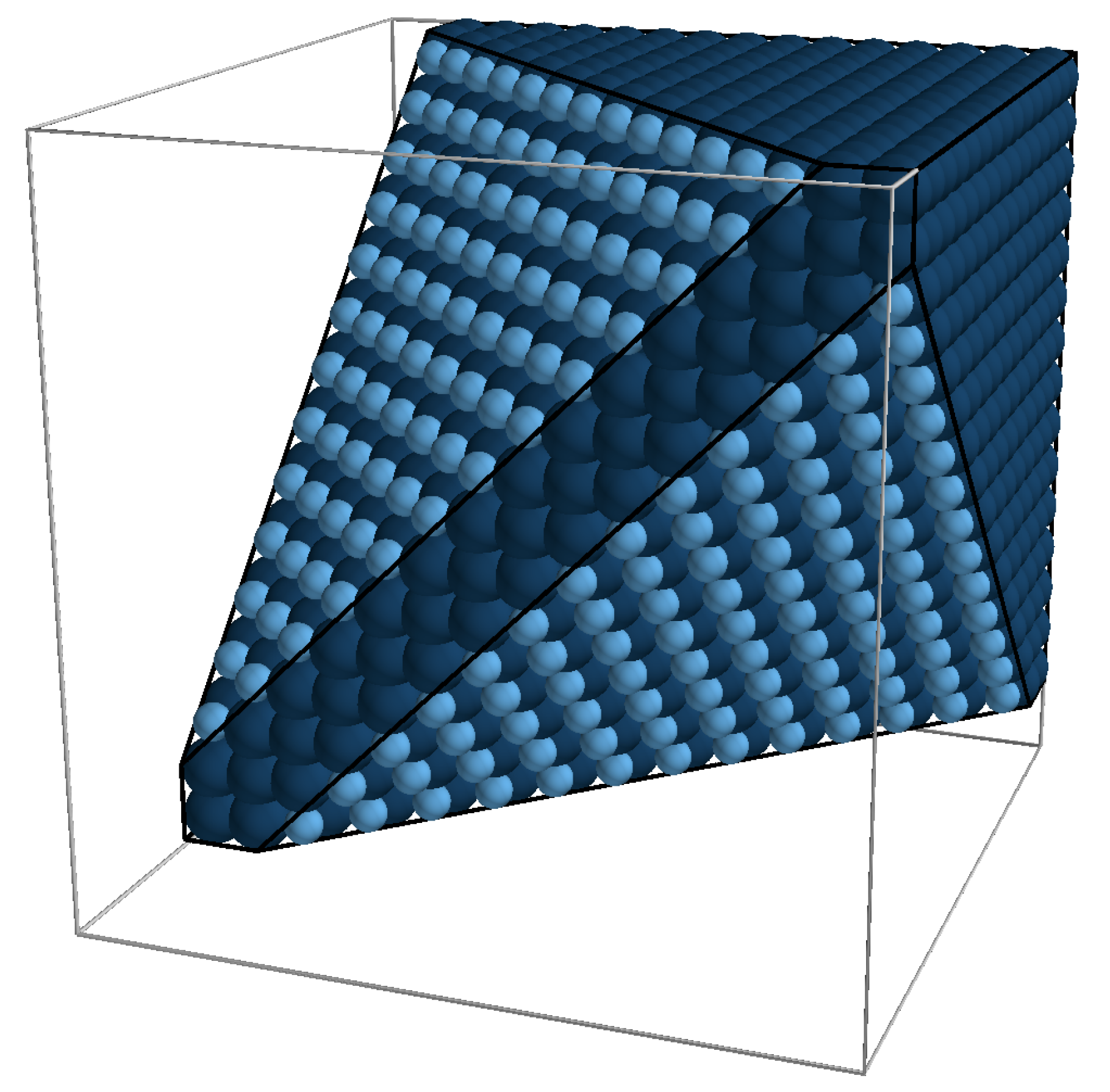}
    \quad
    \includegraphics[width=0.4\textwidth]{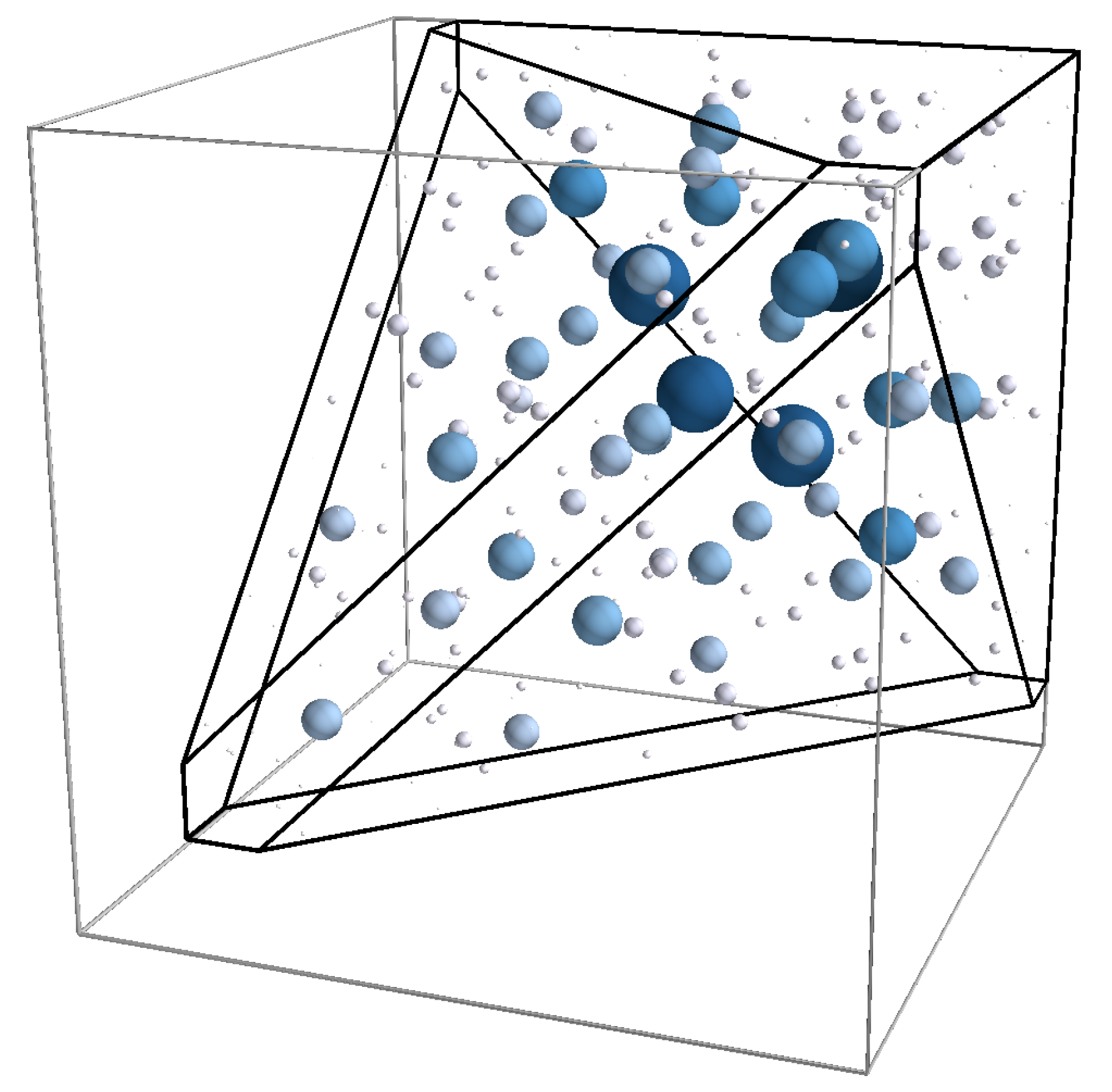}
    \caption{Three-dimensional visualization of the numerical quadrature rules on the tetrapyd domain $\tetra(0.1,1)$ (black edges) inside a unit cube (grey edges). The position and the relative size of each sphere correspond to the given quadrature's evaluation point and its weight, respectively. Shown on the left side is the `uniform' quadrature with 2517 points which directly generalizes the trapezoidal rule. On the right side is obtained using \tetraquad\; ($N=15$, $M=40$), which achieves $\sim$2 orders of magnitude higher precision with a smaller subset of 302 evaluation points.}
    \label{fig:tetraquad_hires}
\end{figure}

Figure \ref{fig:tetraquad_hires} visualizes two quadrature rules over $\tetra(0.1,1)$: the simple uniform quadrature (left) and our new tetrapyd quadrature (right). The spheres are located at the points $\vv{x}_n$ in \eqref{eqn:quadrature_definition} where the integrands are evaluated, and their sizes are proportional to weights $w_n$. Larger spheres are also painted with darker colours for better visualization. As expected, the uniform quadrature has equally sized spheres except near the planes that enforce the triangle inequalities, which cut down the voxel volumes and reduce the weights. Starting with $N=15$ points on each dimension, the uniform quadrature amounts to 2517 grid points in total, or 519 with symmetry. On the other hand, the tetrapyd quadrature only has 302 points or 72 with symmetry. Note that more central points in the tetrapyd tend to have larger weights. Both quadratures respect the symmetry enjoyed by $\tetra$ ($S_3$).

\subsection{Numerical validation}

\begin{figure}
    \includegraphics[width=\textwidth]{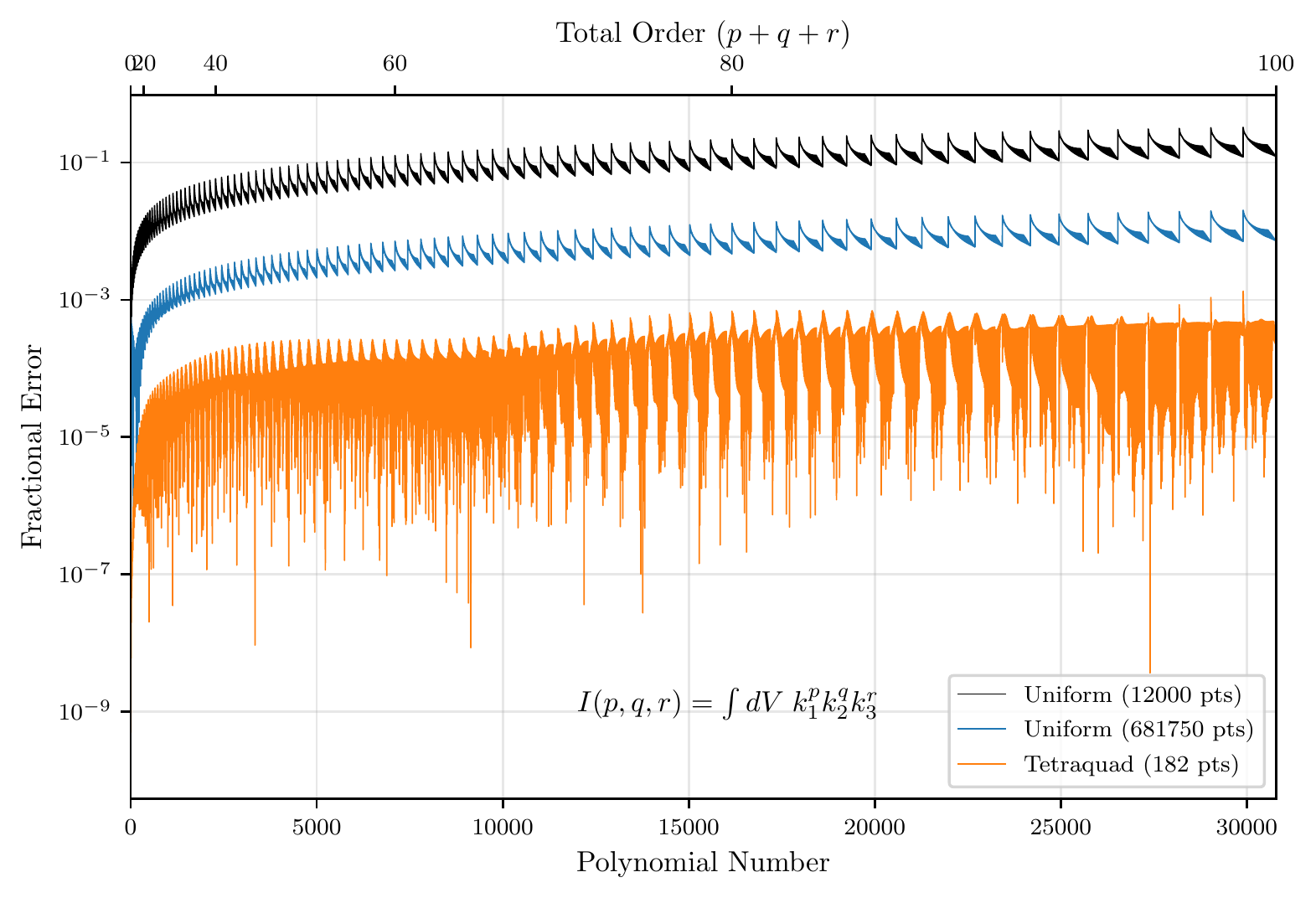}
    \caption{Numerical accuracy of our quadrature rule on integrating $f(k_1,k_2,k_3)=k_1^p k_2^q k_3^r$ over the tetrapyd $\tetra(0.001,1)$. We compare the result with the simple three-dimensional trapezoidal rule on a uniform grid (`Uniform'). Our quadrature (`Tetraquad') achieves much lower fractional error with significantly fewer grid points and has fractional error $<10^{-3}$ for polynomials with total orders (${p+q+r}$) up to 100.}
    \label{fig:tetraquad_total_order}
\end{figure}

We use our public code \tetraquad\; to obtain numerical quadrature rules for $\tetra(0.001,1)$ and test how accurately they can evaluate the integral of $k_1^p k_2^q k_3^r$. For comparison, we create a simple quadrature (labelled as `uniform') with a uniformly spaced grid and weights that are proportional to their voxel volumes within tetrapyd, as described earlier. \footnote{In practice, we use a Monte Carlo method to compute the volumes of voxels that are sliced by the surfaces of tetrapyd; $100,000$ sample points are drawn uniformly from the cube and we count the fraction of samples that lie inside the tetrapyd.} The results are shown in Figure \ref{fig:tetraquad_total_order}.

We find that the \tetraquad\; quadrature with 182 points yields orders of magnitude better precision compared to the simple uniform quadrature with 681,750 points (200 on each axis, restricted to $\tetrasix$). This quadrature is obtained by minimising the error of polynomials of total orders up to 50, but we see that the error remains small ($<10^{-3}$) for total orders up to 100. Note that the number of polynomials $k_1^p k_2^q k_3^r$ of total order $d$ is equal to the nearest integer to $(d+6)^2/12$, \footnote{This is equal to the number of partitions of $n+3$ into exactly 3 parts, or equivalently, the number of partitions where the maximum partition size exactly equal to 3. The proof is given in \cite{hardy1920partition}.} so there are $\sim d^3/36$ linearly independent polynomials with total order less than equal to $d$. Our 182 points quadrature remains accurate for more than $30,000$ independent polynomials.

\begin{figure}
    \includegraphics[width=\textwidth]{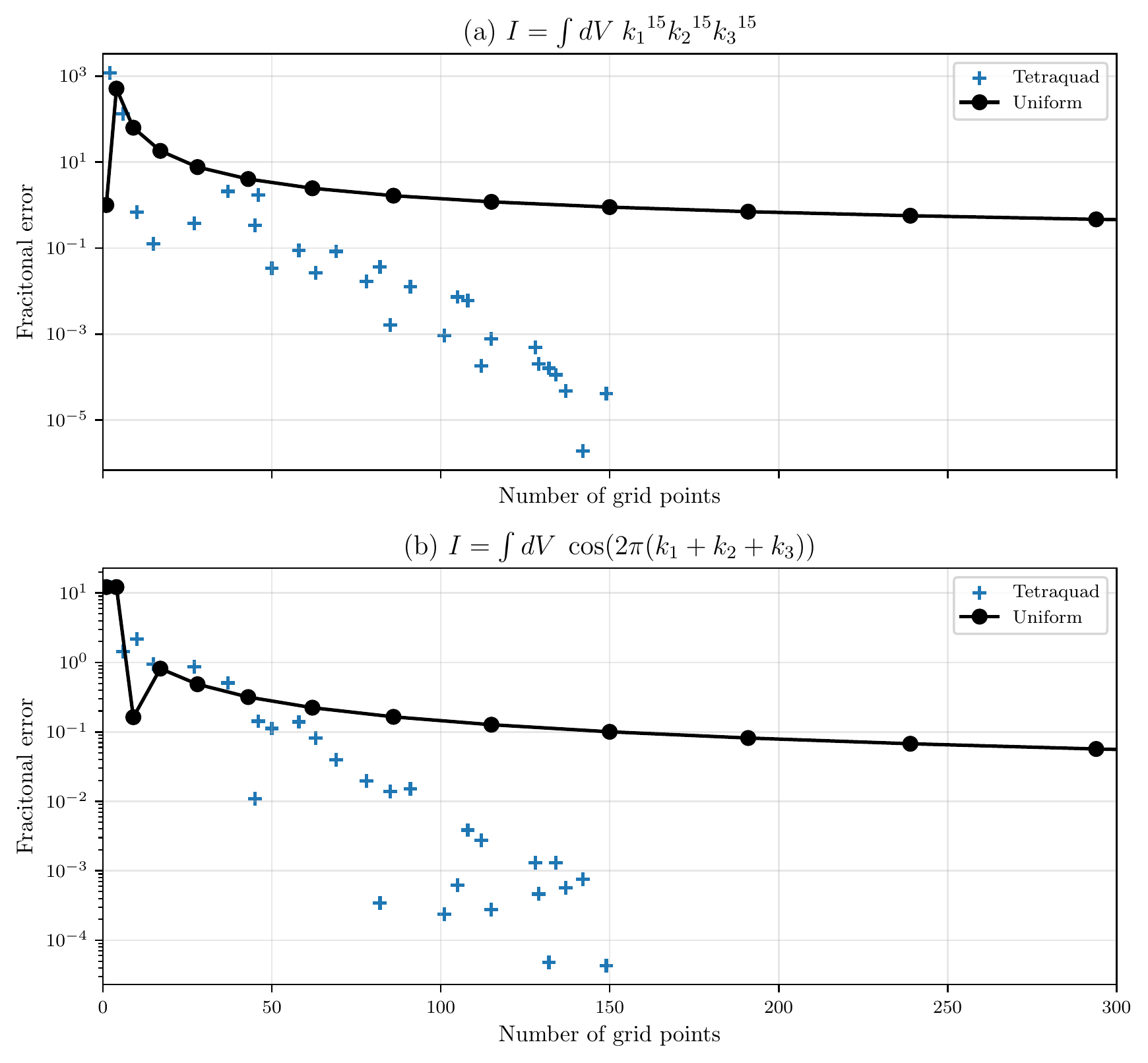}
    \caption{Numerical accuracy of our quadrature rules with varying grid sizes on integrating two test functions over the tetrapyd $\tetra(0.001,1)$. We compare the result with the simple three-dimensional trapezoidal rule on a uniform grid (`Uniform') as before. For both test functions, we see that our quadrature rule performs and scales better; its precision is several orders of magnitude better with the same number of grid points. Note that analytical forms of the two integrals were used to compute the errors.}
    \label{fig:tetraquad_varying_size}
\end{figure}

Next, we investigate how the accuracy of our quadrature scales with the number of grid points. Figure \ref{fig:tetraquad_varying_size} shows how the fractional errors of integrating $k_1^{15} k_2^{15} k_3^{15}$ and $\cos(2\pi (k_1 + k_2 + k_3))$ depend on the number of grid points. The latter was chosen to test how robust the quadrature is when applied to non-polynomial functions. An analytic expression for the integrals of $\sin(\omega (k_1+k_2+k_3) + \phi)$ over tetrapyd was used to evaluate the numerical accuracy.

For grid sizes less than 50, Tetraquad shows comparable or sometimes worse performance compared to the uniform quadrature of similar grid size. However, it improves much faster with the grid size and becomes a few orders of magnitude more precise than the uniform quadrature by $100+$ grid points for both of the test functions. It is also worth noting that Tetraquad remains accurate for functions that are not polynomials.

\section{Results and discussion}       \label{section:results}

We implemented \best\ using C and Python independently from other existing pipelines. Details about the implementation and high-performance computing optimization can be found in Appendix \ref{section:appendix_implementation}.

\subsection{Public release of the code}     \label{section:public_release}

We release the public code for obtaining the Planck 2018 CMB constraints on given bispectrum shapes of interest\textemdash effectively the CMB bispectrum likelihood\textemdash as a Python package \bestcode, which can be found at 
\begin{align}
    \mathrm{\texttt{https://github.com/Wuhyun/CMB-BEST}} \nonumber
\end{align}
\noindent together with the installation guidelines and an example notebook. The code requires a set of precomputed data for some basis functions. Provided with the code is an HDF5 data file that contains the precomputed results for the Planck 2018 CMB temperature and polarization dataset under the monomials and Legendre basis functions.

\vspace{10pt}

\textit{Description --} Speaking in the language of the \best\, formalism, \bestcode\, contains the Cython code for the basis expansion given in \eqref{eqn:BEST_basis_expansion} to compute the expansion coefficients $\alpha$ from a given bispectrum shape. The data file provided contains the precomputed values of $\beta^\mathrm{cub}$ in \eqref{def:beta_cubic}, $\beta^\mathrm{lin}$ in \eqref{def:beta_linear}, and $\Gamma$ in \eqref{def:gamma_dr_integrals}. The $\fNL$ estimate then follows from simple matrix multiplications given by \eqref{eqn:bispectrum_estimator_alpha_beta}. The error $\sigma(\fNL)$ is computed from the sample variance of the $\fNL$ estimates for Gaussian simulations. The Fisher errors in \eqref{eqn:bispectrum_estimator_alpha_beta} are also computed for reference.

Planck 2018 best-fit parameters \cite{PlanckCollaboration2018} were used to compute the CMB transfer functions, with the exception of the scalar amplitude $A_\mathrm{s}$ and tilt $n_\mathrm{s}$, which can be varied in the code. Planck 2018 CMB map from SMICA component separation \cite{PlanckCollaboration2018component}, together with the 160 FFP10 end-to-end simulated maps \cite{PlanckCollaboration2015simulations,PlanckCollaboration2018hfi} with Gaussian initial conditions were used.

\vspace{10pt}

\textit{Outputs --} \bestcode\ outputs the $\fNL$ constraints for the given set of models as $\fNL^{(i)} \pm \sigma(\fNL^{(i)})$, together with the expansion coefficients $\alpha^{(i)}$ and the Fisher matrix $F_{ij}$. If the Fisher matrix is invertible, the marginalized constraints from a joint analysis are provided as well. The package provides some utility functions to save the results in various formats including \texttt{pandas} dataframes and \LaTeX\ tables.

\vspace{10pt}

\textit{Conventions --} Our conventions follow the Planck PNG conventions closely \cite{PlanckCollaboration2013,PlanckCollaboration2015,PlanckCollaboration2018}. The `Model' instance of \bestcode\ is specified by the shape function $S_\Phi$ defined as
\begin{align}
    S_\Phi (k_1, k_2, k_3) \equiv (k_1 k_2 k_3)^2 B_\Phi (k_1,k_2,k_3),
\end{align}
where the primordial potential bispectrum $B_\Phi$ satisfies
\begin{align}
    \langle \Phi(\mathbf{k}_1) \Phi(\mathbf{k}_2) \Phi(\mathbf{k}_3) \rangle = (2\pi)^3 \delta^{(3)}(\mathbf{k}_1 + \mathbf{k}_2 + \mathbf{k}_3) B_\Phi (k_1, k_2, k_3).
\end{align}
The Fourier convention with $\int dx$ and $\int dk/(2\pi)$ is used throughout this work. Under this convention, the local shape function is given by (reiterating \eqref{eqn:local_shape_function} but with a non-unit $n_\mathrm{s}$) 
\begin{align}
    S_\Phi^\mathrm{local}(k_1, k_2, k_3) &\equiv 6 A^2 \; \frac{1}{3}\left[ \frac{k_1^2}{k_2^{2-n_\mathrm{s}} k_3^{2-n_\mathrm{s}}} + \frac{k_2^2}{k_3^{2-n_\mathrm{s}} k_1^{2-n_\mathrm{s}}} + \frac{k_3^2}{k_1^{2-n_\mathrm{s}} k_2^{2-n_\mathrm{s}}}  \right], \label{eqn:local_shape_function_re}
\end{align}
where the power spectrum amplitude $A$ follows the convention in \cite{PlanckCollaboration2018}; $P_\Phi(k) = A k^{\ns - 4}$. Note that this is different but closely related to the scalar power spectrum amplitude $\As$ that appears in the power spectrum analyses such as \cite{PlanckCollaboration2018} via $A = 2\pi^2 \left( \frac{3}{5} \right)^2 k_*^{1 - \ns} \As$, where the pivot scale $k_* = 0.05\mathrm{Mpc}^{-1}$. A detailed comparison between these somewhat confusing conventions is given in Appendix \ref{appendix:pps_amplitude_conventions}. Lastly, note that the shape functions are often normalised with a factor of $6A^2$ so that, in $\zeta$ space, $S_\zeta(k_*,k_*,k_*)=\frac{9}{10} (2\pi)^4 \As^2$.

\vspace{10pt}

\textit{Caveats --} The CMB bispectrum constraints obtained from \bestcode\ are as accurate as the basis expansion. While we have extensively tested that our basis set can accurately handle most classes of models studied in Planck, we advise the users to keep an eye on the convergence statistics included in the resulting constraints. The `convergence correlation' is defined as ${r \equiv \langle S, S^\mathrm{proj} \rangle / \sqrt{\langle S, S \rangle \langle S^\mathrm{proj}, S^\mathrm{proj} \rangle }}$ and measures the `cosine' between the original and projected (basis-expanded) shape functions. This value should be close to 1. The `convergence epsilon' ${\epsilon \equiv \sqrt{2(1-r^2)}}$ is a rough estimate of the expected level of the offset in $\fNL$ caused by the inaccurate expansion, as discussed in the Appendix of \cite{PlanckCollaboration2013}. We currently provide two different resolutions for the Legendre basis: $p_\mathrm{max} = 10$ and $p_\mathrm{max} = 30$. For most smooth shape functions, the standard $p_\mathrm{max}=10$ basis should be sufficient. The higher resolution setting with $p_\mathrm{max}=30$ is desired for more oscillatory shape functions. A rule of thumb is that the Legendre basis would struggle to decompose a shape function with more than $p_\mathrm{max}$ oscillations in the $k$ range of $\left[ 2\times 10^{-4}, 2\times 10^{-1} \right]$.

\subsection{Planck 2018 constraints}    \label{section:results_planck_2018}

\best\ have been thoroughly tested on internal consistency and validated against Planck's Modal estimator. In this section, we reproduce some results from the Planck 2018 analysis \cite{PlanckCollaboration2018} on the standard templates: local, equilateral and orthogonal. We note that the analyses shown in this part are mostly not new and are based heavily on \cite{PlanckCollaboration2013,PlanckCollaboration2015,PlanckCollaboration2018}. The following part (Section \ref{section:results_targeted_analysis}) shows some new constraints on highly oscillatory templates not studied by Planck.

Comparison between the constraints from \best\ and those of Planck is shown in Table \ref{table:standard_constraints}. Both the monomials basis set \eqref{eqn:monomials_mode_functions}, which is equivalent to Planck's KSW estimator, and the Legendre basis \eqref{eqn:legendre_mode_functions} have been used for consistency checks. All results shown assume the $\Lambda$CDM baseline cosmology with the background parameters fixed to the best-fit values for Planck 2018 power spectrum TTTEEE+lowE+lensing \cite{PlanckCollaboration2018Parameters}, although it has been shown that the bispectrum constraints are not very sensitive to these choices \cite{PlanckCollaboration2015}. Following Planck, we use the multipole range of $2\le \ell \le 2500$ for temperature and $4\le \ell \le 2000$ for E-mode polarization.

\begin{table}[htbp!]
    \caption{Comparison between $\fNL$ constraints for the standard templates by \best\ and Planck using the Planck 2018 SMICA maps. The Planck results are quoted from \cite{PlanckCollaboration2018}. For \best, the monomials \eqref{eqn:monomials_mode_functions} and Legendre basis \eqref{eqn:legendre_mode_functions} have been used, while for Planck, the KSW, Binned, Modal 2 \cite{Fergusson2014} estimator results are shown. The results are based on independent single-shape analyses with the lensing-ISW bias subtracted and errors corresponding to 68\% confidence levels. SMICA foreground-cleaned maps have been used.}
    \label{table:standard_constraints}
    \renewcommand{\arraystretch}{1.5} 
    \begin{tabular}{m{0.12\textwidth}m{0.15\textwidth}m{0.10\textwidth}m{0.10\textwidth}<{\centering}m{0.10\textwidth}<{\centering}}
    \toprule
               &             &       &      \multicolumn{2}{c}{$\fNL \pm \sigma(\fNL)$} \\
    \cmidrule{4-5}
    Template & Analysis & Method &    T  &       T+E     \\
    \midrule
    Local & CMB-BEST & KSW &    $-1.7 \pm 6.0$ &    $-1.1 \pm 5.1$ \\
               &             & Legendre &    $-1.4 \pm 6.4$ &    $-1.1 \pm 5.3$ \\
    \cmidrule{2-5}
               & Planck 2018 & KSW &    $-0.5 \pm 5.6$ &    $-0.9 \pm 5.1$ \\
               &             & Binned &  $-0.1 \pm 5.6$ &    $-2.5 \pm 5.0$ \\
               &             & Modal &  $-0.6 \pm 6.4$ &    $-2.0 \pm 5.0$ \\
    \midrule
    Equilateral & CMB-BEST & KSW &   $14 \pm 66$ &  $-22 \pm 49$ \\
               &             & Legendre &   $15 \pm 66$ &  $-22 \pm 49$ \\
    \cmidrule{2-5}
               & Planck 2018 & KSW   &   $7 \pm 66$ &   $-18 \pm 47$ \\
               &             & Binned &   $26 \pm 69$ &   $-19 \pm 48$ \\
               &             & Modal &   $34 \pm 67$ &   $-4 \pm 43$ \\
    \midrule
    Orthogonal & CMB-BEST & KSW &   $-9 \pm 40$ &  $-31 \pm 24$ \\
               &             & Legendre &   $-9 \pm 40$ &  $-32 \pm 24$ \\
    \cmidrule{2-5}
               & Planck 2018 & KSW   &  $-15 \pm 36$ &  $-37 \pm 23$ \\
               &             & Binned&  $-11 \pm 39$ &  $-34 \pm 24$ \\
               &             & Modal &  $-26 \pm 43$ &  $-40 \pm 24$ \\
    \bottomrule
    \end{tabular}
\end{table}

We find that the constraints from \best\ are consistent with Planck and the differences are within the scatter between Planck's own estimators, which is explained in the appendix of \cite{PlanckCollaboration2013}. The results between the two basis sets of \best\ are entirely consistent as well. The small differences come from the fact that the Legendre basis functions have a limited $k$ range, which slightly reduces the power at very low $\ell$s. This affects the local template the most by reducing the power of squeezed configurations (e.g. $b_{2\ell\ell}$) and increases the error bars by up to 6\%. For further checks, we performed a map-by-map validation of \best\ against the Modal estimator 160 simulated maps with Gaussian initial conditions, which is detailed in Appendix \ref{section:appendix_map_by_map}. We found the results from the two independent methods to be consistent on these simulated maps. Note also that the mean squared error (MSE) in the basis expansion for each of these models is kept less than $10^{-8}$ in all cases.

The constraints can be understood as hypothesis tests for the individual templates with $\hat{f}_\mathrm{NL}$ as a test statistic: testing the alternative hypothesis $H_1 : \, \fNL \neq 0$ against the null hypothesis $H_0 : \, \fNL = 0$. The $\fNL$ estimates of the 160 simulated maps follow a Gaussian distribution with zero mean under the null hypothesis. If the $p$-value of the observation $\hat{f}_\mathrm{NL}$ with respect to this distribution is sufficiently low (e.g. $<0.01$), \footnote{In principle, the $p$-value threshold should be set before seeing the data.} we would reject the null hypothesis. As can be seen from Table \ref{table:standard_constraints}, we do not have sufficient evidence to reject the null hypotheses of $\fNL=0$ for the three standard templates.

The constraints above are from independent single-shape analyses; for each template and the corresponding $B^\mathrm{temp}_\ells$, we consider the model $B^\mathrm{th}_\ells =\fNL B^\mathrm{temp}_\ells $. \textit{If} the model is correct and the template is an accurate representation of the primordial bispectrum, then the result is an estimate of the amplitude of PNG in the given shape. Various inflationary models that predict different amounts of $\fNL$ can be constrained by this estimate.

{Alternatively, we could do a joint analysis where our theoretical bispectrum is given by a linear combination of either all or some of the templates under consideration. Here, for demonstration purposes, we assume that a model predicts a bispectrum shape that is represented as a linear combination of all three standard templates: ${ B^\mathrm{th}_\ells = \fNL^\mathrm{local} B^\mathrm{local}_\ells + \fNL^\mathrm{equil} B^\mathrm{equil}_\ells + \fNL^\mathrm{ortho} B^\mathrm{ortho}_\ells }$. The $\fNL$ estimate is given by \eqref{eqn:cmb_bispectrum_estimator_MLE} in this case. For templates that are orthogonal in shape (with respect to the inner product \eqref{eqn:tetrapyd_inner_product}), the Fisher matrix tends to be diagonal, and the constraints are nearly identical to those of single-shape analyses. Figure \ref{fig:standard_templates_correlation} shows correlations between the three standard templates, both in the primordial space with the inner product \eqref{eqn:tetrapyd_inner_product} and in the $\fNL$ estimates. The templates are weakly correlated as expected, with some noticeable correlation between the local and orthogonal templates.

\begin{figure*}[htbp!] 
	\centering    
	\includegraphics{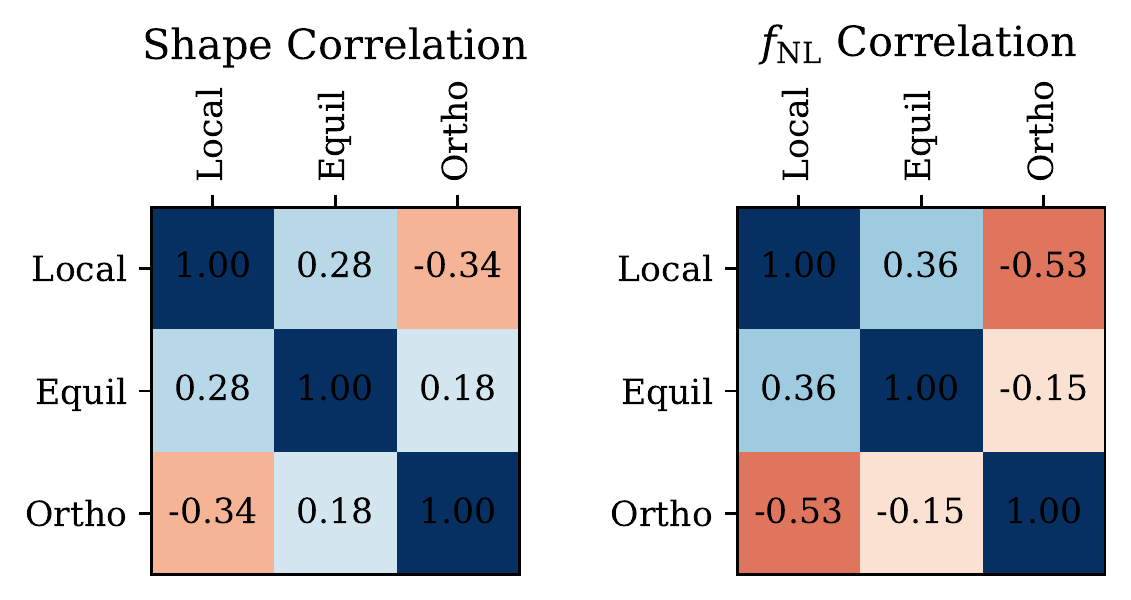}
	\caption{Correlation matrix for the standard templates: Local, Equilateral and Orthogonal. On the left is the correlation (`cosine') between templates in the primordial $(k_1,k_2,k_3)$ space, while on the right is the correlation between single $\fNL$ estimates obtained from 160 simulated maps. Both temperature and polarization maps were used for the latter. We find that the standard templates are mostly orthogonal to each other in both the primordial and late-time spaces, with the exception of local and orthogonal which are more correlated late-time. }
	\label{fig:standard_templates_correlation}
\end{figure*}

A joint analysis was performed to obtain the likelihood $\mathcal{L}(\mathbf{B}^\mathrm{th} | \hat{\mathbf{B}}) = \mathcal{L}( \fNL^\mathrm{local},  \fNL^\mathrm{equil},  \fNL^\mathrm{ortho} | \hat{\mathbf{B}})$ as defined in \eqref{eqn:bispectrum_likelihood_formal}. The marginalized likelihood of the three $\fNL$ parameters is shown in Figure \ref{fig:triangle_marginal_fisher_TP}, plotted using the Python library \texttt{GetDist} \cite{Lewis2019getdist}. Overall, there is no evidence for a non-zero $\fNL$.

\begin{figure*}[htbp!] 
	\centering    
	\includegraphics{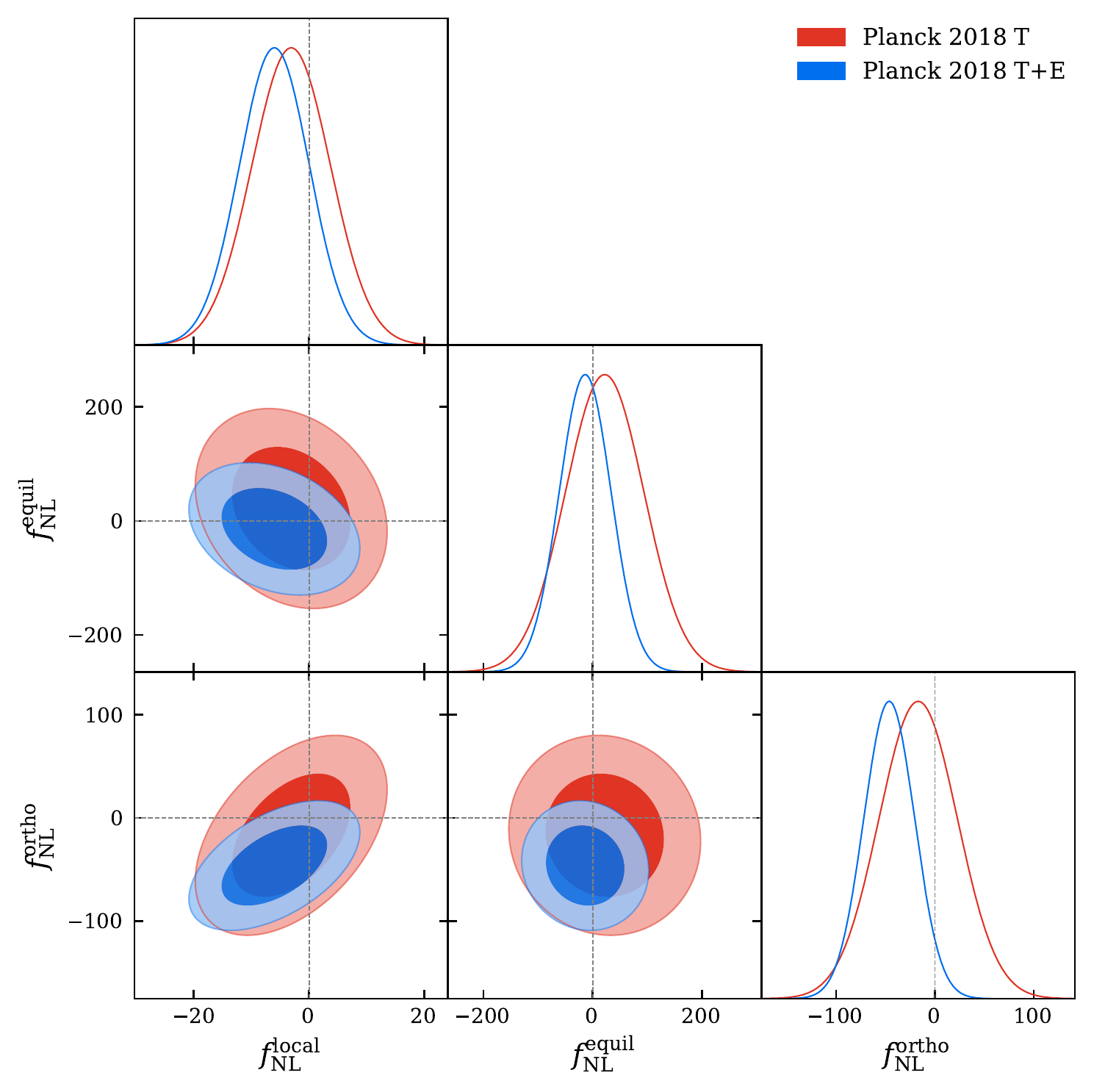}
	\caption{The CMB bispectrum likelihood for a joint analysis on the local, equilateral and orthogonal templates using \best. The contours and line plots show marginalized likelihoods obtained from the temperature-only (T) and temperature + polarization (T+E) data. Note that the Fisher matrix was used in the likelihood, which is an accurate description with only a slight underestimation of the errors (see Appendix \ref{section:appendix_optimality} for details). The lensing-ISW bias has been taken into account. The dashed lines correspond to $\fNL=0$ in each plot. \best's Legendre basis set was used for this analysis. }
	\label{fig:triangle_marginal_fisher_TP}
\end{figure*}

The likelihood in Figure \ref{fig:triangle_marginal_fisher_TP} is plotted under the assumption that $\{ \hat{B}_\ells \}$ is multivariate normal distributed with diagonal covariances equal to $(6/\Delta_\ells)C_{\ell_1}C_{\ell_2}C_{\ell_3}$, as we discussed when writing down the likelihood in \eqref{eqn:bispectrum_likelihood_formal}. The validity of this assumption for this analysis is tested in Appendix \ref{section:appendix_optimality} using the simulated maps. The Fisher matrix slightly underestimates the error (up to 10\% in orthogonal), but the estimator remains nearly optimal in all cases.

Assuming that the likelihood is an accurate representation of the CMB bispectrum statistics, there can be multiple interpretations of the results depending on one's statistical alignments. A frequentist would interpret the contours as the distribution of $\fNL$ estimates we could have obtained under different realisations of the Universe and independent observations. They would conclude that there is no significant evidence for rejecting the null hypothesis of $\hat{\mathbf{f}}_\mathrm{NL}=\mathbf{0}$. When testing an early universe model that predicts $\mathbf{f}_\mathrm{NL}=\mathbf{x}$ for some $\mathbf{x}$, the likelihood can similarly be used to construct a $p$-value test for a rejection/non-rejection of the model.

On the other hand, a Bayesian-minded may interpret the contours as the posterior distributions for $\mathbf{f}_\mathrm{NL}$ when the prior is flat and wide. The posterior distribution represents the probability distribution of each $\fNL$ \textit{given} the model and the data. Alternatively, the constraints from the Large-Scale Structure surveys can be used as priors. As the current constraints from LSS are weaker in comparison, the posteriors would appear similar to the plots in Figure \ref{fig:triangle_marginal_fisher_TP}. The Bayesian model comparison methods such as Akaike information criteria (AIC) or Bayes factors could be used to compare different models from various templates against the model without PNG. Note that unfortunately the constant term appearing in \eqref{eqn:fNL_likelihood} is not computed directly in \best, KSW or Modal estimator, so the likelihood is determined only up to a constant multiplicative factor.

\subsection{Targeted oscillatory basis}     \label{section:results_targeted_analysis}

Having validated the pipeline internally and against Planck, we present some new constraints on $\fNL$ for templates that have not been studied before in Planck analysis \cite{PlanckCollaboration2018}. Motivated by various inflationary models which predicts linearly-spaced oscillations in the bispectrum, we consider the bispectrum shapes of form
\begin{align}
    (k_1 k_2 k_3)^2 B(k_1, k_2, k_3) = f(k_1, k_2, k_3) \sin(\omega (k_1 + k_2 + k_3) + \phi),
\end{align}
where $f(k_1,k_2,k_3)$ is some envelope function of choice and $\omega$ and $\phi$ parametrizes the overall linearly-spaced oscillations. 

The case where $f(k_1,k_2,k_3)=1$ was thoroughly studied using a KSW-like estimator \cite{Munchmeyer2014,PlanckCollaboration2018} up to oscillation frequencies $0 \le \omega \le 3000$ in units of Mpc. This was viable due to the separability of linear oscillations. However, such a trick does not immediately apply to a more complex type of envelope function $f$, and it is challenging to go to a highly oscillatory regime; the polynomial basis of the Modal estimator often lacks resolution, and the oscillations are washed out by binning in the Binned estimator. Planck's Modal studied general oscillatory templates up to $\omega \lesssim 350$ Mpc.

Using the flexibility of mode function choices in \best, we can construct a targeted basis with a fixed value of $\omega$ as shown in \eqref{eqn:sine_legendre_mode_functions}. This allows us to study general shape functions $f(k_1,k_2,k_3)$ with a fixed oscillation frequency $\omega$. As a part of the proof-of-concept, we performed a targeted analysis on $\omega=1000$ Mpc. This frequency was chosen somewhat arbitrarily for demonstration purposes, but in future analyses, noticeable oscillatory signals found in the power spectrum or bispectrum could guide this choice. The results are summarized in Table \ref{table:targeted_constraints}. Overall, we find no significant evidence from the CMB bispectrum for non-zero PNG in these templates. 

\begin{table}[htbp!]
    \caption{Constraints on $\fNL$ for highly oscillatory bispectrum templates highly oscillatory templates obtained using \best's targeted basis. Templates are products of the standard templates with linear oscillations: $S(k_1,k_2,k_3)=f(k_1,k_2,k_3)\sin(\omega (k_1+k_2+k_3) + \phi)$ for $\omega=1000$ Mpc, where $f$ is the local, equilateral, or orthogonal template. The constant and exponential envelopes are defined in the text. Overall, we do not find evidence for PNG with oscillations with $\omega=1000$ Mpc. }
    \label{table:targeted_constraints}
    \renewcommand{\arraystretch}{1.5} 
    \begin{tabular}{m{0.12\textwidth}m{0.05\textwidth}m{0.15\textwidth}<{\centering}m{0.10\textwidth}<{\centering}}
    \toprule
           Shape & Phase &  $\fNL \pm \sigma(\fNL)$ &  $\fNL/\sigma(\fNL)$ \\
    \midrule
            Local & sin &        $-0.36 \pm  0.97$ & -0.37 \\
                  & cos &        $-1.36 \pm 1.10$ & -1.24 \\
      Equilateral & sin &         $0.28 \pm  0.32$ &  0.89 \\
                  & cos &         $0.45 \pm  0.45$ &  0.99 \\
       Orthogonal & sin &         $0.10 \pm  0.12$ &  0.90 \\
                  & cos &         $0.15 \pm   0.16$ &  0.93 \\
         Constant & sin &       $-18 \pm 14$ & -1.27 \\
                  & cos &        $10 \pm 15$ &  0.69 \\
    Exponential     & sin &       $-15 \pm 26$ & -0.56 \\
                  & cos &       $-16 \pm 24$ & -0.67 \\
    \bottomrule
    \end{tabular}
\end{table}

The templates are normalized so that the envelope part $f(k_1,k_2,k_3)$ follows the conventions from Planck \cite{PlanckCollaboration2013}. The `constant' envelope is equivalent to the constant feature models studied in Planck \cite{PlanckCollaboration2018}. For the `exponential' envelope, we set $f(k_1,k_2,k_3)\propto \exp \left[ -((1/3)(k_1+k_2+k_3) - k_*)^2/2 d^2 \right]$, where $k_*=0.05\;\mathrm{Mpc}^{-1}$ and $d=0.02\;\mathrm{Mpc}^{-1}$ were chosen as an example. All constraints shown are independent single-shape analyses.

In order to test the ability of the \best\ to differentiate these oscillatory templates, we plot the correlation matrix between the $\fNL$ estimates obtained from 160 Gaussian simulations. The results are shown in Figure \ref{fig:sinleg_correlations}.

\begin{figure}
    \centering
    \includegraphics{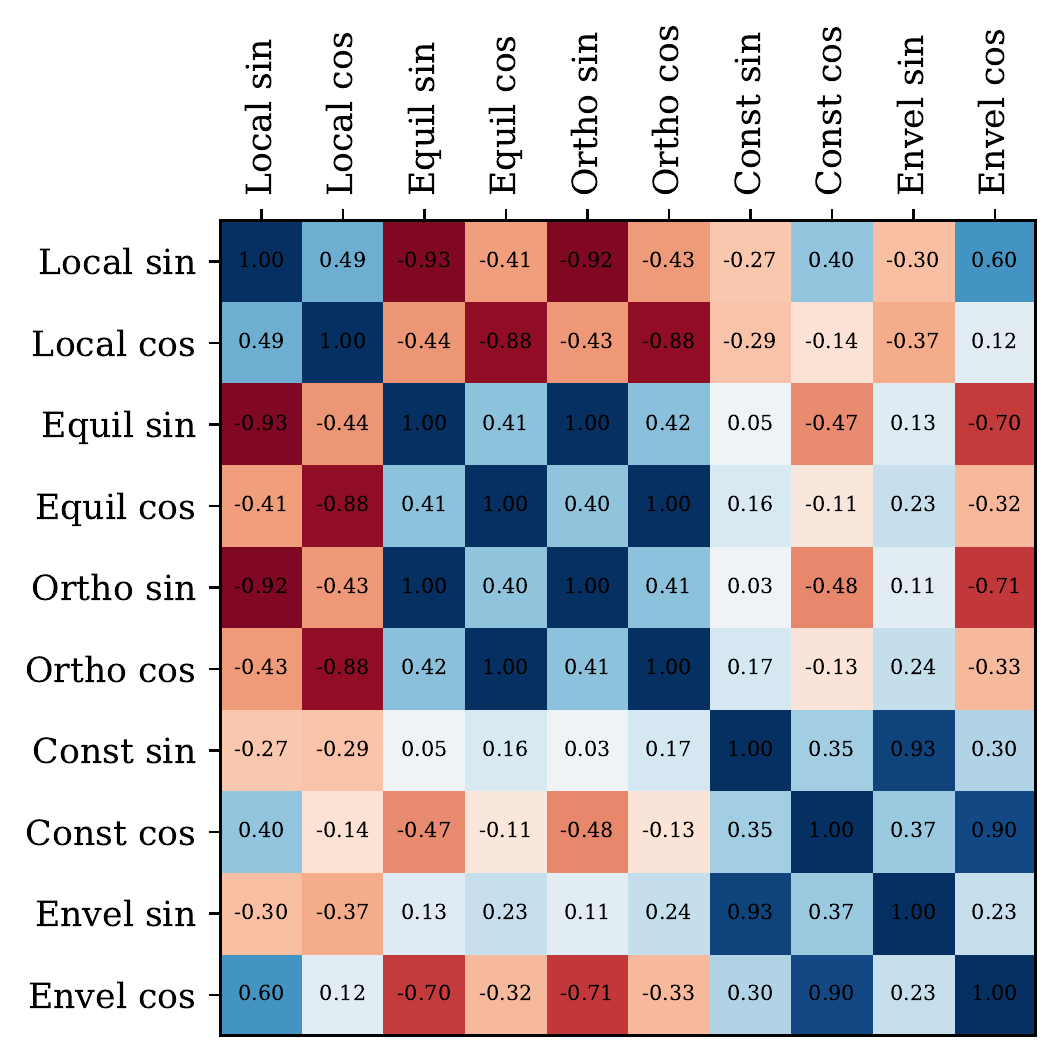}
    \caption{Correlation matrix for highly oscillatory bispectrum templates obtained using \best's targeted basis. The models considered are the same as in Table \ref{table:targeted_constraints} and are described in the text.  }
    \label{fig:sinleg_correlations}
\end{figure}

We find that the sine and cosine templates with out-of-phase oscillations always remain relatively uncorrelated, while within the same phases of oscillations, the local, equilateral and orthogonal templates all show strong correlations ($\ge 0.88$) in the $\fNL$ values they retrieve from the simulations. The in-phase templates are as uncorrelated in the primordial space as the correlation between the envelopes shown in Figure \ref{fig:standard_templates_correlation} since our basis expansion for the envelope is accurate (MSE less than $10^{-8}$). The differences, however, appear to get washed out by the projection effects; the templates predict nearly identical CMB bispectrum despite having different envelopes.

The constant and exponential envelopes are found to be mostly uncorrelated with the templates with local, equilateral and orthogonal envelopes. The two of them are highly correlated themselves, which suggests that most of the bispectrum signatures from oscillations come from the enveloped region around $K=0.15\; \mathrm{Mpc}^{-1}$.

A joint analysis between these enveloped shapes can cover models which predict different oscillation phases at different limits. For example, considering the local and equilateral envelopes only, the best-fit bispectrum for the 4-parameter model (3 up to normalization) is given by:
\begin{align}
    B^\mathrm{bf}(k_1,k_2,k_3) = 5.46\, B^\mathrm{Local}(k_1,k_2,k_3) \sin(\omega K + 38^{\circ}) + 1.46\, B^\mathrm{Equil}(k_1,k_2,k_3) \sin(\omega K + 77^{\circ}),
\end{align}
where $K\equiv k_1+k_2+k_3$ and $\omega=1000\mathrm{Mpc}$. Note that oscillations have different phases in the equilateral and squeezed limits. A single-shape analysis on this template gives the constraint $4.2 \pm 2.1$ with a $2.0\sigma$-level significance, which is not very high especially given that we had 3 free parameters to fit for. We conclude that there is no significant evidence for these highly oscillatory ($\omega=1000\mathrm{Mpc}$) templates with the Planck data.

\section{Conclusion} \label{section:conclusion}

In this work, we presented the formalism and implementation of our new high-resolution bispectrum estimator \best. We publicly release the frontend of our code as \bestcode, together with a data file containing pre-computed results from the computation-heavy parts of the pipeline, so that researchers can test get CMB bispectrum constraints for their theoretical bispectrum predictions using our code.

\best\ is formulated and coded for general choices of mode functions and can be used for various purposes. When using a small set of monomials as mode functions, the code is equivalent to the KSW estimator and gives quick constraints on the standard bispectrum templates. Using a larger set of Legendre polynomials allows an extensive analysis covering a broad range of models. Lastly, a targeted oscillatory basis enables an in-depth, high-resolution analysis of models with a specific oscillation frequency.

We have thoroughly optimized the code, tested the consistency of the results, and showed a proof-of-concept example of a high-resolution bispectrum analysis. In doing so, we came up with some methodological advances in algorithmic optimization and numerical computation. In particular, \texttt{tetraquad} is a code for computing numerical quadrature rules for integrating functions over a tetrapyd domain. The quadrature rule is efficient, accurate and completely general, and is expected to benefit various bispectrum analyses in CMB and LSS. 

We plan on studying various inflationary models using this code, and extend the methodology and code to existing and future CMB data, including the upcoming Simons observatory.

\newpage

\begin{acknowledgments}

WS would like to thank Philip Clarke for all his support throughout the development and validation process of \best, Arman Shafieloo, David Parkinson and Rodrigo Calderon for many valuable discussions, and Kacper Kornet for the high-performance computing support. WS was supported by the project 우주거대구조를 이용한 암흑우주 연구 (Understanding Dark Universe Using Large Scale Structure of the Universe) funded by the Ministry of Science. WS was also supported by Internal Graduate Studentship from Trinity College, Cambridge during his PhD when a part of this work was undertaken.

\end{acknowledgments}

\appendix

\section{Spherical harmonics related identities} \label{section:appendix_identities}

In this section, we quote useful identities related to spherical harmonics that appear in  the bispectrum formalism. The equations are quoted from various sources including \cite{Dodelson2003textbook,Fergusson2020cosmology,Komatsu2010,Sohn2022thesis}.

The Gaunt integral is defined as an integral of spherical harmonics $Y_{\ell m}$ over the sphere $S^2$:
\begin{align}
    \mathcal{G}^\ells_{m_1 m_2 m_3} \equiv \int d^2 \hat{\vv{n}} Y_{\ell_1 m_1}(\hat{\vv{n}}) Y_{\ell_2 m_2}(\hat{\vv{n}}) Y_{\ell_3 m_3}(\hat{\vv{n}}).
\end{align}
It can be rewritten in terms of the Wigner 3j symbols as follows.
\begin{align}
    \mathcal{G}^\ells_{m_1 m_2 m_3} &= \sqrt{ \frac{(2\ell_1+1)(2\ell_2+1)(2\ell_3+1)}{4\pi} } \begin{pmatrix} \ell_1& \ell_2& \ell_3 \\ m_1& m_2& m_3 \end{pmatrix} \begin{pmatrix} \ell_1& \ell_2& \ell_3 \\ 0& 0& 0 \end{pmatrix}  \\
    &= \begin{pmatrix} \ell_1& \ell_2& \ell_3 \\ m_1& m_2& m_3 \end{pmatrix} h_\ells.  \label{eqn:gaunt_in_terms_of_h}
\end{align}
The Wigner 3j symbols are normalized so that
\begin{align}
     \sum_{m_j} \begin{pmatrix} \ell_1& \ell_2& \ell_3 \\ m_1& m_2& m_3 \end{pmatrix}^2 = 1,
\end{align}
and therefore
\begin{align}
    \sum_{m_j} \begin{pmatrix} \ell_1& \ell_2& \ell_3 \\ m_1& m_2& m_3 \end{pmatrix} \mathcal{G}^\ells_{m_1 m_2 m_3} &= h_\ells,    \\
    \sum_{m_j} \left( \mathcal{G}^\ells_{m_1 m_2 m_3} \right)^2 &= h_\ells^2,
\end{align}
as long as $(\ell_1,\ell_2,\ell_3)$ form a triangle.

When $m_1=m_2=m_3=0$, a useful identity relates Wigner 3j symbols with Legendre polynomials:
\begin{align}
    \begin{pmatrix} \ell_1& \ell_2& \ell_3 \\ 0& 0& 0 \end{pmatrix}^2 = \frac{1}{2} \int_{-1}^{1} d\mu \; P_{\ell_1}(\mu) P_{\ell_2}(\mu) P_{\ell_3}(\mu). 
\end{align}
It immediately follows that
\begin{align}
    h^2_\ells = \frac{2\ell_1+1)(2\ell_2+1)(2\ell_3+1)}{8\pi} \int_{-1}^{1} d\mu \; P_{\ell_1}(\mu) P_{\ell_2}(\mu) P_{\ell_3}(\mu). \label{eqn:h2_in_terms_of_legendre_polys}
\end{align}

\section{Analytic formula for integrals of polynomials over a general tetrapyd} \label{section:appendix_analytic_tetrapyd_formula}

In this section, we write down the analytic formula for the integral of polynomials over a general tetrapyd defined as
\begin{align}
    \tetra(\alpha,1) \equiv \{(k_1,k_2,k_3) : 2\max \{k_1, k_2, k_3\} \le k_1 + k_2 + k_3 \;\mathrm{and}\; \alpha \le k_1, k_2, k_3 \le 1 \},
\end{align}
for $0 \le \alpha \le 1/2$. Note that the integral over ${\tetra(\kmin,\kmax)}$ can be derived from that of ${\tetra(\kmin/\kmax,1)}$ We have derived the formula given here by hand with occasional aid from Mathematica \cite{Mathematica}.

Instead of the tetrapyd with rather complicated shape, we consider a cube $[\alpha,1]^3$ and the volumes inside this cube excluded by the tetrapyd:
\begin{align}
    V_\mathrm{cube} &= \tetra(\alpha,1) \cup V_1 \cup V_2 \cup V_3, \label{eqn:general_tetrapyd_volume_split} \\
    V_1 &\equiv \{(k_1,k_2,k_3) : k_2 + k_3 \le k_1 \;\mathrm{and}\; \alpha \le k_2, k_3 \le k_1 \le 1 \}, \\
    V_2 &\equiv \{(k_1,k_2,k_3) : k_3 + k_1 \le k_2 \;\mathrm{and}\; \alpha \le k_3, k_1 \le k_2 \le 1 \}, \\
    V_3 &\equiv \{(k_1,k_2,k_3) : k_1 + k_2 \le k_3 \;\mathrm{and}\; \alpha \le k_3, k_1 \le k_3 \le 1 \}.
\end{align}
Since each set on the right hand side of \eqref{eqn:general_tetrapyd_volume_split} is disjoint, we have
\begin{align}
    \int_{\tetra(\alpha,1)} dV \; k_1^{p} k_2^{q} k_3^{r} = I_\mathrm{cube} - I_1 - I_2 - I_3.
\end{align}
The integral over the cube is straightforward due to separability;
\begin{align}
    I_\mathrm{cube} = \int_{ [\alpha,1]^3} dV \; k_1^{p} k_2^{q} k_3^{r} = \frac{(1-\alpha^{p+1})(1-\alpha^{q+1})(1-\alpha^{r+1})}{(p+1)(q+1)(r+1)}.
\end{align}
On the other hand, 
\begin{align}
    I_1 =\;& \int_{V_1} dk_1 dk_2 dk_3 \; k_1^p k_2^q k_3^r \\
    =\;& \int_{\alpha}^{1-\alpha} dk_2 \int_{\alpha}^{1-k_2} dk_3 \int_{k_2+k_3}^{1} dk_1 \; k_1^p k_2^q k_3^r  \\
    =\;& \frac{1}{p+1} \int_{\alpha}^{1-\alpha} dk_2 \int_{\alpha}^{1-k_2} dk_3 \; k_2^q k_3^r \left[ 1 - (k_2 + k_3)^{p+1} \right] \\
    \vdots\;& \nonumber \\
    =\;& \frac{1}{p+1} \int_{2\alpha}^{1} ds \; s^{q+r+1}(1-s^{p+1}) \; \mathrm{Beta}(\frac{\alpha}{s},1-\frac{\alpha}{s};q+1,r+1) \\
    \vdots\;& \nonumber \\
    =\;& \frac{\mathrm{Beta}(\alpha,1-\alpha; q+1, r+1)}{(p+q+r+3)(q+r+2)} \nonumber \\
    &- \frac{1}{(p+1)(q+1)(r+1)} \left[ \frac{(r+1)\alpha^{r+1}(1-\alpha)^{q+1}}{q+r+2} + \frac{(q+1)\alpha^{q+1}(1-\alpha)^{r+1}}{q+r+2} - \alpha^{q+r+2} \right] \nonumber \\
    &+ \frac{\alpha^{p+q+r+3}}{(p+1)(p+q+r+3)} \left[(-1)^q \mathrm{Beta}(2,\frac{1}{\alpha}; p+2, q+1) + (-1)^r \mathrm{Beta}(2,\frac{1}{\alpha}; p+2, r+1) \right].
\end{align}
Above, we used a change of variables ${(k_2, k_3) = (st, s(1-t))}$ and integration by parts with ${du = s^{q+r+1}(1-s^{p+1})ds}$ and ${v=\mathrm{Beta}(\frac{\alpha}{s},1-\frac{\alpha}{s};q+1,r+1})$.  The incomplete beta function is defined as
\begin{align}
    \mathrm{Beta}(x, y; p, q) \equiv \int_{x}^{y} dt \; t^{p-1} (1-t)^{q-1},
\end{align}
so that $\mathrm{Beta}(0,1;p,q)$ corresponds to the usual (complete) beta function. The integrals $I_2$ and $I_3$ are obtained similarly by cycling $(p,q,r)$ around. Although the definition originally restricts $x$ and $y$ to $[0,1]$, the function can be analytically continued to the region outside. We compute these by using the ordinary hypergeometric function ${}_2F_1 (a,b;c;z)$ and its relation to the incomplete beta function \cite{DLMF};
\begin{align}
    \mathrm{Beta}(x, y; p, q) = \frac{y^p (1-y)^ q}{p} {\;}_2F_1 (p+q,1;p+1;y) - \frac{x^p (1-x)^ q}{p} {\;}_2F_1 (p+q,1;p+1;x).
\end{align}
In our public code \texttt{Tetraquad}, the hypergeometric function is computed using \texttt{mpmath} \cite{mpmath}, a Python library for arbitrary-precision arithmetic. Note that the formula above holds from general $p,q,r$ that are not necessarily integers.

\section{Bispectrum estimator including polarization} \label{section:appendix_full_cmb_bispectrum_estimator}

In this section, we generalise the CMB bispectrum likelihood \eqref{eqn:bispectrum_likelihood_formal} to include both CMB temperature and E-mode polarization.

For simplicity, we assume that the covariance matrix is diagonal:
\begin{align}
    \left< a^{X_1}_{\ell_1 m_1} a^{X_2}_{\ell_2 m_2} \right> = C^{X_1 X_2}_{\ell_1} \delta_{\ell_1 \ell_2} \delta_{m_1 -m_2},
\end{align}
where $X=\mathrm{T},\mathrm{E}$ denotes temperature and E-mode polarization, respectively.

The CMB bispectrum now consists of multiple parts: TTT, TTE, TEE and EEE. The bispectrum estimate \eqref{eqn:observed_bispectrum} can be generalised as
\begin{align}
    \hat{B}^{X_1 X_2 X_3}_\ells &\equiv \sum_{m_j} \begin{pmatrix} \ell_1& \ell_2& \ell_3 \\ m_1& m_2& m_3 \end{pmatrix} \Bigl[  a^{X_1}_{\ell_1 m_1} a^{X_2}_{\ell_2 m_2} a^{X_3}_{\ell_3 m_3} - \left[ \langle a^{X_1}_{\ell_1 m_1} a^{X_2}_{\ell_2 m_2} \rangle a^{X_3}_{\ell_3 m_3} + (2~\mathrm{cyc.})  \right] \Bigr]. \label{eqn:observed_bispectrum_polarisation} \\
    &= \sum_{m_j} \begin{pmatrix} \ell_1& \ell_2& \ell_3 \\ m_1& m_2& m_3 \end{pmatrix} \left(B^{X_1 X_2 X_3}\right)^{\ells}_{m_1 m_2 m_3}.
\end{align}

As discussed in \cite{Fergusson2014,Sohn2019}, it is convenient to work with linear combinations of $X=\mathrm{T},\mathrm{E}$ in which the covariance matrix $C_\ell^{X_1 X_2}$ at each multipole $\ell$ is diagonal. Considering the Cholesky decomposition of $C_\ell^{-1} = L_\ell^\mathrm{T} L_\ell$, where
\begin{align}
    L_\ell \equiv \begin{pmatrix} \frac{1}{\sqrt{C_\ell^\mathrm{TT}}} & 0  \\ \frac{- C_\ell^\mathrm{TE}}{\sqrt{C_\ell^\mathrm{TT}} \sqrt{C_\ell^\mathrm{TT} C_\ell^\mathrm{EE} - \left( C_\ell^\mathrm{TE} \right)^2}}  &  \frac{ C_\ell^\mathrm{TT}}{\sqrt{C_\ell^\mathrm{TT}} \sqrt{C_\ell^\mathrm{TT} C_\ell^\mathrm{EE} - \left( C_\ell^\mathrm{TE} \right)^2}} \end{pmatrix}.
\end{align}
The $C_\ell$s here include the corrections from the beam and noise. We now work with $\tilde{X}=\tilde{\mathrm{T}},\tilde{\mathrm{E}}$ obtained by transforming $X=\mathrm{T},\mathrm{E}$ through $L$. The CMB transfer function  $T_\ell(k)$ and observed $a_{\ell m}$s transform as
\begin{align}
    T_\ell^{\tilde{X}} (k) = \sum_{X} L_\ell^{\tilde{X} X} T_\ell^{X}, \quad\text{and}\quad a_{\ell m}^{\tilde{X}} = \sum_{X} L_\ell^{\tilde{X} X} a_{\ell m}^{X}.
\end{align}
It follows that $\left< a^{\tilde{X}_1}_{\ell_1 m_1} a^{\tilde{X}_2}_{\ell_2 m_2} \right> = \delta_{\ell_1 \ell_2} \delta_{m_1 -m_2}$.

Analogously to the temperature-only analysis, we assume that $\{ \hat{B}_\ells \}_{\ell_1 \le \ell_2 \le \ell_3}$ have a diagonal covariance matrix with diagonal entries  $(6 / \Delta_\ells) ~ C_{\ell_1} C_{\ell_2} C_{\ell_3}$, but now with $2\times 2$ matrices $C_\ell^{X_1 X_2}$. Under this assumption, the CMB bispectrum likelihood can be written as 
\begin{align}
    \mathcal{L}(\mathbf{B}^\mathrm{th} | \hat{\mathbf{B}}) \propto \mathrm{exp} \Biggl[ -&\frac{1}{2} \sum_{\ell_1 \le \ell_2 \le \ell_3} \sum_{X_j, X'_j} \biggl[ \frac{ \Delta_\ells }{6} \left( \hat{B}^{X_1 X_2 X_3}_\ells - B^{\mathrm{th},X_1 X_2 X_3}_\ells \right) \nonumber \\
    &\left( C_{\ell_1}^{-1} \right)^{X_1 X_1'} \left( C_{\ell_2}^{-1} \right)^{X_2 X_2'} \left( C_{\ell_3}^{-1} \right)^{X_3 X_3'}  \left( \hat{B}^{X'_1 X'_2 X'_3}_\ells - B^{\mathrm{th},X'_1 X'_2 X'_3}_\ells \right) \biggr] \Biggr].  \label{eqn:bispectrum_likelihood_formal_pol}
\end{align}

After the transformation from $X$ to $\tilde{X}$, the likelihood \eqref{eqn:bispectrum_likelihood_formal_pol} simplifies to
\begin{align}
    \mathcal{L}(\mathbf{B}^\mathrm{th} | \hat{\mathbf{B}}) \propto \mathrm{exp} \Biggl[ -&\frac{1}{2} \sum_{\ell_1 \le \ell_2 \le \ell_3} \sum_{\tilde{X}_j} \frac{ \Delta_\ells }{6} \left( \hat{B}^{\tilde{X}_1 \tilde{X}_2 \tilde{X}_3}_\ells - B^{\mathrm{th},\tilde{X}_1 \tilde{X}_2 \tilde{X}_3}_\ells \right)^2  \Biggr].
\end{align}

Given a set of bispectrum templates and their amplitude parametrized by $\fNL$,
\begin{align}
    B^{\mathrm{th},X_1 X_2 X_3}_\ells = h_\ells \sum_i \fNL^{(i)}  b^{(i),X_1 X_2 X_3}_\ells.
\end{align}
The maximum likelihood estimator for $\fNL$ is then given by
\begin{align}
    \widehat{\fNL}^{(i)} = \sum_j (F^{-1})_{ij} S_j,
\end{align}
where
\begin{align}
    S_i &\equiv \frac{1}{6} \sum_{\ell_j, m_j, \tilde{X}_j} \mathcal{G}^\ells_{m_1 m_2 m_3} b^{(i),\tilde{X}_1 \tilde{X}_2 \tilde{X}_3}_\ells \Bigl[ a^{\tilde{X}_1}_{\ell_1 m_1} a^{\tilde{X}_2}_{\ell_2 m_2} a^{\tilde{X}_3}_{\ell_3 m_3} - \left[ \langle a^{\tilde{X}_1}_{\ell_1 m_1} a^{\tilde{X}_2}_{\ell_2 m_2} \rangle a^{\tilde{X}_3}_{\ell_3 m_3} + (2~\mathrm{cyc.})  \right] \Bigr],  \label{eqn:cmb_bispectrum_MLE_signal_pol} \\
    F_{ij} &\equiv \frac{1}{6} \sum_{\ell_j, \tilde{X}_j} h^2_\ells b^{(i),\tilde{X}_1 \tilde{X}_2 \tilde{X}_3}_\ells b^{(j),\tilde{X}_1 \tilde{X}_2 \tilde{X}_3}_\ells .  \label{eqn:cmb_bispectrum_MLE_fisher_pol}
\end{align}

\vspace{10pt}
The \best\ formalism can effectively exploit the separability of the summation over $\tilde{X}$. The projected modes defined in \eqref{def:projected_modes} are extended to polarization:
\begin{align}
    \tilde{q}^{\tilde{X}}_{p}(\ell, r) \equiv \frac{2r^\frac{2}{3}}{\pi} \int dk \; q_p(k) T^{\tilde{X}}_\ell (k) j_\ell (kr). 
\end{align}
The filtered maps \eqref{def:filtered_maps} for the T+E analyses has a simple summation over $\tilde{X}$ as follows.
\begin{align}
    M^{(i)}_p (\hat{\vv{n}}, r) \equiv \sum_{\ell,m,\tilde{X}} \tilde{q}^{\tilde{X}}_p (l,r) a^{\tilde{X}}_{\ell m} Y_{\ell m} (\hat{\vv{n}}).
\end{align}
The rest of the \best\ formalism is identical to those described in the main text. Therefore, adding polarization does not significantly affect the computational complexity; it only doubles the number of spherical harmonic transforms (SHT) needed to obtain the filtered maps.

\section{Primordial power spectrum amplitude conventions} \label{appendix:pps_amplitude_conventions}

Here, we clarify and relate the different notations used in a) Planck 2018 constraints on PNG \cite{PlanckCollaboration2018}, b) Planck 2018 cosmological parameters \cite{PlanckCollaboration2018Parameters}, and c) Xingang Chen's review on PNG \cite{Chen2010review} for the primordial power spectrum amplitudes.

In \cite{PlanckCollaboration2018}, the primordial power spectrum $P_\Phi(k)$ is defined as
\begin{align}
    P_\Phi (k) = A k^{\ns - 4},
\end{align}
so that
\begin{align}
    \langle \Phi(\vv{k}_1) \Phi(\vv{k}_2) \rangle = P_\Phi(k_1) (2\pi)^3 \delta^{(3)}(\vv{k}_1 + \vv{k}_2).
\end{align}
This is the \textit{dimensionful} primordial power spectrum. Note that $A$ is written as $\Delta_\Phi$ in \cite{Fergusson2012}.

The \textit{dimensionless} primordial power spectrum $\mathcal{P}_\zeta$ (denoted $P_\zeta$ in \cite{Chen2010review}) is defined from
\begin{align}
    \langle \Phi(\vv{k}_1) \Phi(\vv{k}_2) \rangle = \frac{\mathcal{P}_\Phi(k_1)}{2k_1^3} (2\pi)^5 \delta^{(3)}(\vv{k}_1 + \vv{k}_2),
\end{align}
so that
\begin{align}
    \mathcal{P}_\Phi(k) = \frac{k^3}{2\pi^2} P_\Phi(k).
\end{align}

At superhorizon scales, the Bardeen potential $\Phi$ and the curvature perturbation $\zeta$ are related via $\Phi=(3/5)\zeta$. Hence,
\begin{align}
    P_\Phi(k) = \left( \frac{3}{5} \right)^2 P_\zeta(k), \quad \mathcal{P}_\Phi(k) = \left( \frac{3}{5} \right)^2 \mathcal{P}_\zeta(k).
\end{align}

In \cite{PlanckCollaboration2018Parameters}, the scalar amplitude $\As$ and the spectral index $\ns$ are defined through the \textit{dimensionless} curvature power spectrum:
\begin{align}
    \mathcal{P}_\zeta(k) = \As \left( \frac{k}{k_*} \right)^{\ns - 1},
\end{align}
for some pivot scale $k_*=0.05 \mathrm{Mpc}^{-1}$ so that $\mathcal{P}(k_*)=\As$. This $\As$ is identical to $\tilde{P}_\zeta$ in \cite{Chen2010review}.

Putting these altogether, we relate $A$ from \cite{PlanckCollaboration2018} and $\As$ from \cite{PlanckCollaboration2018Parameters} as follows:
\begin{align}
    A k^{\ns -4} = P_\Phi(k) = \frac{2\pi^2}{k^3} \left( \frac{3}{5} \right)^2 \mathcal{P}_\zeta(k) = \frac{2\pi^2}{k^3} \left( \frac{3}{5} \right)^2 \As \left( \frac{k}{k_*} \right)^{\ns - 1},
\end{align}
and therefore
\begin{align}
    A = 2\pi^2 \left( \frac{3}{5} \right)^2 k_*^{1 - \ns} \As.
\end{align}
Above is the final conversion relation between the amplitude Planck PNG paper's \cite{PlanckCollaboration2018} amplitude $A$ and the Planck parameter estimation paper \cite{PlanckCollaboration2018Parameters}.

\section{Implementation and optimisation} \label{section:appendix_implementation}

The \best\ formalism significantly reduces the computational complexity of CMB bispectrum estimation, but obtaining the linear term $\beta^\text{lin}$ in (\ref{def:beta_linear}) can still be prohibitively expensive unless thoroughly optimised. Better performance also means higher resolution, as we may include mode functions in our basis.

Here, we detail our optimisation process: algorithm design, parallel computing, and data locality improvements. Throughout this section, the tensors and functions are treated as discrete multi-dimensional arrays. Our notation for indices and their limits are summarised in Table \ref{table:index_conventions}. A simple trapezoidal rule is used for most numerical integrals except the $\mu$ integral in (\ref{def:gamma_dr_integrals}), where the Gauss-Legendre quadrature computed from the public code \texttt{Quadpts} \cite{Hale2013} is used for the highly oscillatory integrand. Multi-dimensional arrays are stored in the row-major order following the \textsc{C} convention. The \texttt{Healpix} library \cite{Gorski2005healpix} is used for pixelisation of the sky, and its component library \texttt{Libsharp} \cite{Reinecke2013libsharp} for the spherical harmonic transforms (SHTs). Note that a large portion of the material in this appendix is adapted from \cite{Sohn2022thesis}.

\begin{table*}[htbp]
	\caption{Our index conventions for discretised arrays and their range.}
	\centering
	\label{table:index_conventions}
	\renewcommand{\arraystretch}{1.5} 
	\begin{tabular}{m{0.1\textwidth}  m{0.1\textwidth}  m{0.7\textwidth}} \toprule
		Index & Range & Description \\
		
		\midrule
		$r$ & $[0, N_r)$ & Line-of-sight integral $r$ grid index. \\
		
		$p, p_j$ & $[0, p_\mathrm{max})$ & Mode number. $p_j$ is a shorthand for $(p_1, p_2, p_3)$. \\
		
		$i,j$ & $[0, N_\mathrm{sim}]$ & Map number. Index $i=0$ corresponds to the observed map, while $i>0$ corresponds to different FFP10 simulated maps \cite{PlanckCollaboration2015simulations}. \\			
		
		$n$ & $[0, N_\mathrm{pix})$ & Map pixel number. \\ 
		
		$\ell,m$ & $[\ell_\mathrm{min}, \ell_\mathrm{max}]$ & Spherical harmonic multipole moments. Note $-\ell \le m \le \ell$. \\
		
		$\mu$ & $[0, N_\mu)$ & Grid index for the Gauss-Legendre quadrature. \\
		
		\bottomrule
	\end{tabular}
\end{table*}

\subsection{Algorithm}

There are three key quantities to be precomputed and stored for \best: $\Gamma$ \eqref{def:gamma_dr_integrals}, $\beta^\text{cub}$ \eqref{def:beta_cubic}, and $\beta^\mathrm{lin}$ \eqref{def:beta_linear}. The matrix $\Gamma$ directly relates to the Fisher information matrix for the estimator, while the two $\beta$'s provide the amplitude of $\fNL$ for each of the CMB observations and simulations.

In most cases, the bottleneck point of our pipeline is computing the linear term $\beta^\mathrm{lin}$. Even though the $\Gamma$ matrix scales more rapidly with the number of basis functions ($\varpropto p_\mathrm{max}^6$) than the $\beta$'s ($\propto p_\mathrm{max}^3$), it does not involve operations with high-resolution maps and remains subdominant in terms of the total cost in the regime we consider ($p_\mathrm{max} \le 30$).

The discretised versions of \eqref{def:beta_cubic} and \eqref{def:beta_linear} are given by
\begin{align}
	\beta^\text{cub}(i, p_1,p_2,p_3) &= \sum_r \sum_n \; M(r, i, p_1, n) \cdot M(r, i, p_2, n) \cdot M(r, i, p_3, n), \\
	\beta^\text{lin}(i, p_1,p_2,p_3) &= \sum_r \sum_{j \neq i} \sum_n \; M(r, j, p_1, n) \cdot M(r, j, p_2, n) \cdot M(r, i, p_3, n),
\end{align}
respectively. The order of indices is chosen such that the following calculations have optimal memory layouts. Some integral weights and factors are absorbed into arrays for brevity.

First, note that the data arrays for different values of $r$ are completely independent of each other. This provides us with a natural way to split tasks. We compute and save contributions to $\beta$'s for each $r$ separately and summed over at the end with minimal overhead. Therefore, throughout the rest of this chapter, we assume that $r$ is fixed and drop the $r$ dependence in the descriptions of our algorithms.

The filtered map arrays $M(i,p,n)$ are obtained as follows. A given map $i$ is first transformed into spherical harmonic coefficients $a^{(i)}(\ell,m)$s via SHT. We then compute $\tilde{q}_(p,l) * a(i,\ell,m) / C(\ell)$ from (\ref{def:filtered_maps}) through inverse SHT to synthesise the filtered maps.

As a rough guide to the size of each summation, we typically have $N_\mathrm{sim} = 160$ simulations, $p_\mathrm{max} = 30$ modes, and $N_\mathrm{pix} = 50,331,648$ pixels. \footnote{This value corresponds to $N_\mathrm{side} = 2048$ in Healpix: $N_\mathrm{pix} = 12 N_\mathrm{side}^2$} Considering the fact that one double-precision array of size $\sim 50$ million pixels takes about 400MB of memory space, this is indeed a task for supercomputers.

Finding the optimal algorithm for $\beta$ computation was about reducing the computational complexity while keeping the memory footage in check. Computing and saving some intermediate quantities often reduce the computational complexity but can often be infeasible due to maximum memory restraints. We describe our final algorithm below as Algorithm \ref{alg:beta_third_attempt}. A more detailed reasoning can be found in \cite{Sohn2022thesis}.

\begin{algorithm}[H]
    \caption{Computing $\beta$s: fast and memory efficient}\label{alg:beta_third_attempt}
    \begin{algorithmic}[1] 
        \State Allocate $m(p,n)$ \Comment{Memory $\sim p_\mathrm{max} \cdot N_\mathrm{pix}$}
        \State Allocate $C(p_1,p_2,n)$ \Comment{Memory $\sim p_\mathrm{max}^2 \cdot N_\mathrm{pix}$}
        \\
        \For{each map $i$}
            \For{each mode $p$}
                \Comment{$O(N_\mathrm{sims} \cdot p_\mathrm{max} \cdot N_\mathrm{pix}^{3/2})$}
                \State \textbf{compute} $M(i,p,n)$ by SHT and store in $m(p,n)$
            \EndFor
            \\
            \For{each pair of modes $(p_1,p_2)$}
                \For{each pixel $n$}
                    \Comment{$O(N_\mathrm{sim} \cdot p_\mathrm{max}^2 \cdot N_\mathrm{pix})$}
                    \State $C(p_1,p_2,n) \pluseq m(p_1,n) \cdot m(p_2,n)$
                \EndFor
            \EndFor
        \EndFor
        \Comment{$C(p_1,p_2,n)$ ready}
        \\
        \For{each map $i$}
            \For{each of mode $p$}
                \Comment{$O(N_\mathrm{sim} \cdot p_\mathrm{max} \cdot N_\mathrm{pix}^{3/2})$}
                \State \textbf{compute} $M(i,p,n)$ by SHT and store in $m(p,n)$
            \EndFor
            \\
            \For{each set of modes $(p_1,p_2,p_3)$}
                \For{each pixel $n$}
                    \Comment{$O(N_\mathrm{sim} \cdot p_\mathrm{max}^3 \cdot N_\mathrm{pix})$}
                    \State $\beta^\mathrm{cub}(i, p_1, p_2, p_3) \pluseq m(p_1,n) \cdot m(p_2,n) \cdot m(p_3,n)$
                    \State $\beta^\mathrm{lin}(i, p_1, p_2, p_3) \pluseq C(p_1, p_2, n) \cdot m( p_3, n)$
                \EndFor
            \EndFor
        \EndFor
    \end{algorithmic}
\end{algorithm}

The trick is to reduce the amount of memory required at the cost of doubling the SHTs for computing $M(i,p,n)$s which are subdominant. The first time through, SHT results from each map are used to find $C(p_1,p_2,n)$. After a full loop over maps we have $C(p_1,p_2,n)$ ready, another set of SHTs for each map allows us to obtain the $\beta$'s. Algorithm \ref{alg:beta_third_attempt} requires memory of size $\approx p_\mathrm{max}^2 N_\mathrm{pix}$ and has leading computational complexity of $O(N_\mathrm{sim} p_\mathrm{max}^3 N_\mathrm{pix})$. 

SHTs have a subdominant contribution to the total computation time even after becoming doubled in number. One of the main strengths of Algorithm \ref{alg:beta_third_attempt} is that both the memory and computation time scale linearly with the number of simulations used, $N_\text{sims}$. In the future when a larger number of Gaussian simulations are required to acquire a more accurate estimate of the linear term, it is straightforward to adapt our method accordingly.

\subsection{Parallelism}

In order to fully benefit from modern computer architecture, we introduce parallelism in multiple levels of \best\ for optimal performance on supercomputers.

Following \cite{Jeffers2016intel}, we discuss parallelisation at three different levels. They mostly correspond to nodes, cores, and registers/cache in modern computer clusters. Each level has distinct characteristics which make them ideal for different parallelisation techniques. We make full use of each level in our methodology.

The first level of parallelism relates to dividing the main work into many computation-heavy tasks with limited data communication between them. Since the tasks are largely independent, each of them can be assigned to independent nodes or to the Message Passing Interface (MPI) \cite{Gropp1999MPI} ranks. We exploit this level of parallelism in \best\ by splitting the line-of-sight integration over $r$; almost no data is shared between different $r$s despite the heavy operations within each of them. \best\ scales well with the number of computing units assigned for each $r$ point (or several $r$ points).

The second level of parallelism is for multiple computational subtasks on a single set of data. Most modern supercomputers use multi-core processors. Each core, or processing unit, can run one or more threads, executing instructions independently from each other while sharing memory space. MPI would not be as effective here due to the large amount of data sharing required; ranks would have to either continuously communicate with each other, or store duplicate copies of the data. This type of parallelism is required in the SHTs and map operations of \best. We use Open Multi-Processing (OpenMP) \cite{Dagum1998openmp} for multi-threading in \texttt{C} for this purpose.

The third level of parallelism applies to the identical arithmetic operations applied to multiple data items, ideally adjacent in memory space: Single Instruction, Multiple Data (SIMD). Many procedures in \best\ involve large array operations which benefit from this type of parallelism. We use Advanced Vector Extensions (AVX) supported by Intel processors to exploit this, especially for operations involving large filtered map arrays. In particular, Intel's Xeon Phi series' AVX-512 implementation, where the 512-bit registers hold up to 8 double-precision floating numbers \cite{Jeffers2016intel}, provided a major boost to the computation speed.

\subsection{Data locality}

We implemented Algorithm \ref{alg:beta_third_attempt} and profiled it using the Intel VTune Amplifier. Our program was found to be memory-bound, meaning its speed is limited mainly by the speed of memory access. The CPU speed, rate of the file I/O, and MPI communication all have subdominant contributions in comparison. This is somewhat expected since our method deals with large map arrays. The number of operations on each data element is small compared to the size of data, causing the CPUs to be `starved' for data to work on most of the time. Our final set of optimisation focusses on improving memory access patterns.

CPUs of most modern computers contain a small amount of memory attached to them called \textit{cache}. Recently used data and instructions are stored in cache memory so that reusing them is more efficient; accessing them is much faster than loading from the main, larger memory often shared with other CPUs. A cache is often divided into multiple levels. The smallest and fastest is the L1 cache, which is the first level a CPU checks for data. When the required data is not stored in the L1 cache, a \textit{cache miss} occurs. The system then has to look further down the cache levels to fetch the desired data, incurring a large time loss. As the cache `caches' memory locations in units of cache lines, or chunks of memory containing multiple data elements, accessing memory locally significantly increases the chance of cache hits and boosts overall performance.

\best\ has been modified in two ways to improve data locality and memory performance. The first one was simple yet effective; we made sure to initialise large arrays within the same OpenMP construct as the main computation loop. This guarantees the physical memory of array elements to be allocated near the cores where they are going to be used. Memory access during the main computation loop is therefore much faster than it would be otherwise. Systems with non-uniform memory access (NUMA) especially benefit from this method. For \best, we gained a two-times speed-up compared to when a single master thread initialised the entire array.

The second optimisation centres around \textit{cache blocking}, a technique used to maximise data reuse. The most expensive loop Algorithm \ref{alg:beta_third_attempt} is:

\begin{algorithmic}
	\For{each set of modes $(p_1,p_2,p_3)$}
		\For{each pixel $n$}
			\State $\beta^\text{cub}(i, p_1, p_2, p_3) \pluseq m(p_1,n) \cdot m(p_2,n) \cdot m(p_3,n)$
			\State $\beta^\text{lin}(i, p_1, p_2, p_3) \pluseq C(p_1, p_2, n) \cdot m( p_3, n)$
		\EndFor
	\EndFor
\end{algorithmic}

In every outermost loop, four large arrays are read from memory: $m(p_1,\cdot)$, $m(p_2,\cdot)$, $m(p_3,\cdot)$, and $C(p_1, p_2, \cdot)$. Each of them takes up around 400MB of memory. Since their size is greater than the cache storage capacity, data in front of them are gone from the cache by the time a loop over $n$ completes. All four arrays will then have to be loaded from the main memory again when the next iteration starts.

To allow different parts of the array to be reused before they are lost in cache, we divide the large arrays into equally-sized blocks that fit inside the cache memory. We restructure the loop as follows:

\begin{algorithmic}
	\For{each block $b$}
		\For{each set of modes $(p_1,p_2,p_3)$}
			\For{each pixel $n'$ in block}
				\State $\beta^\text{cub}(i, p_1, p_2, p_3) \pluseq m(p_1,n') \cdot m(p_2,n') \cdot m(p_3,n')$
				\State $\beta^\text{lin}(i, p_1, p_2, p_3) \pluseq C(p_1, p_2, n') \cdot m( p_3, n')$
			\EndFor
		\EndFor
	\EndFor
\end{algorithmic}

New pixel numbers are calculated as $n' = B \cdot b + n$, where $B$ is the size of each block and $0 \le n < B$. We have not changed the total number of arithmetic operations required, so the computational complexity remains the same. Meanwhile, data locality within each block is greatly improved, as each of the blocked arrays is now small enough to fit in the cache. Each data element is accessed in closer succession temporarily as well.

One caveat here is that having too many blocks may degrade the overall performance. There exists a non-negligible overhead coming from an extra \textit{for} loop and the OpenMP construct used over $n'$. The block size divided by the number of cores should not be smaller than the size of cache lines either. The optimal size of cache blocks depends on the memory architecture of the processor used. This often needs to be found empirically. For our implementation, we found the optimal number of blocks to be 128, yielding a three-times speed-up compared to the original code without cache blocking.

\section{Validation}    \label{section:appendix_validation}

\subsection{Map-by-map validation of \best}       \label{section:appendix_map_by_map}

In this section, we show map-by-map validation of the \best\ pipeline against the Modal estimator used in Planck analyses.

Figure \ref{fig:map_by_map_Legendre_Monomial_T} and \ref{fig:map_by_map_Legendre_Monomial_TP} show comparisons between the Monomials and Legendre basis sets of \best. A total of 160 Planck's FFP10 SMICA-component-separated simulations \cite{PlanckCollaboration2015simulations,PlanckCollaboration2018component} were used for this analysis. For each simulated map, we compare the $\fNL$s of the standard templates that have been computed using two different bases. The two of them agree almost exactly, with a map-by-map correlation of more than $0.99$ in all cases.

Despite using identical data sets, we see a small scatter between the Monomials and Legendre basis sets in the local $\fNL$. This is mainly due to the limited $k$-range of the Legendre basis: $[2.08\times 10^{-4}, 2.08\times 10^{-1}] \; \mathrm{Mpc}^{-1}$. The range has been set wide enough to cover the scales where CMB information comes from, but also narrow enough to have enough resolution for expanding oscillatory templates with polynomials of maximum order $p_\mathrm{max}$. Loss of large-scale information with $k < 2.08\times 10^{-4}$ has a small effect on equilateral and orthogonal shapes, but affects the local shape through the squeezed limit. The error bar $\sigma(\fNL)$ is about 7\% smaller for the Monomials basis for this reason. Choosing a wider $k$-range for the Legendre basis has been shown to remove this scatter \cite{Sohn2022thesis}.

\begin{figure*}[htbp!] 
	\centering 
	\includegraphics{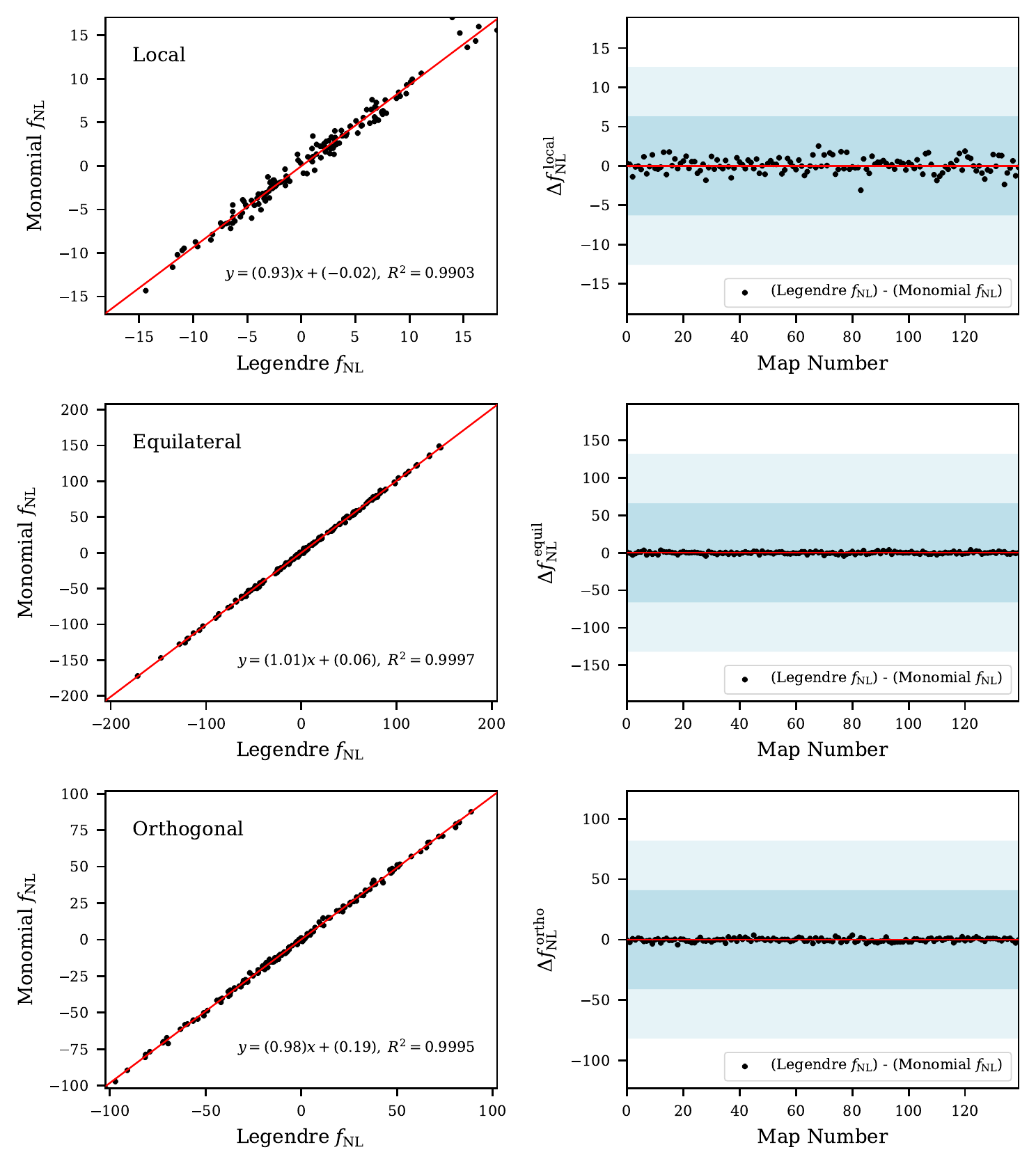}
	\caption{A map-by-map comparison of the $f_\text{NL}$ estimates evaluated using the Monomials (KSW) and Legendre basis sets for three standard templates. The Planck 2018 CMB temperature map and 160 FFP10 simulations have been used, each representing a single point on the scatter plot (left). Details of the linear regression to data (red) are annotated below. Shown on the right-hand side are plots of the differences in the $f_\mathrm{NL}$ values for each map, together with the $1\sigma$ and $2\sigma$ intervals shaded in blue. In the ideal case where the two basis sets yield identical results, we should see all the points lie on the line $y=x$ for the left plot and $y=0$ for the right plot. For more information on each of the three theoretical templates used, see e.g., \cite{PlanckCollaboration2013}.}
    \label{fig:map_by_map_Legendre_Monomial_T}
\end{figure*}

\begin{figure*}[htbp!] 
	\centering 
	\includegraphics{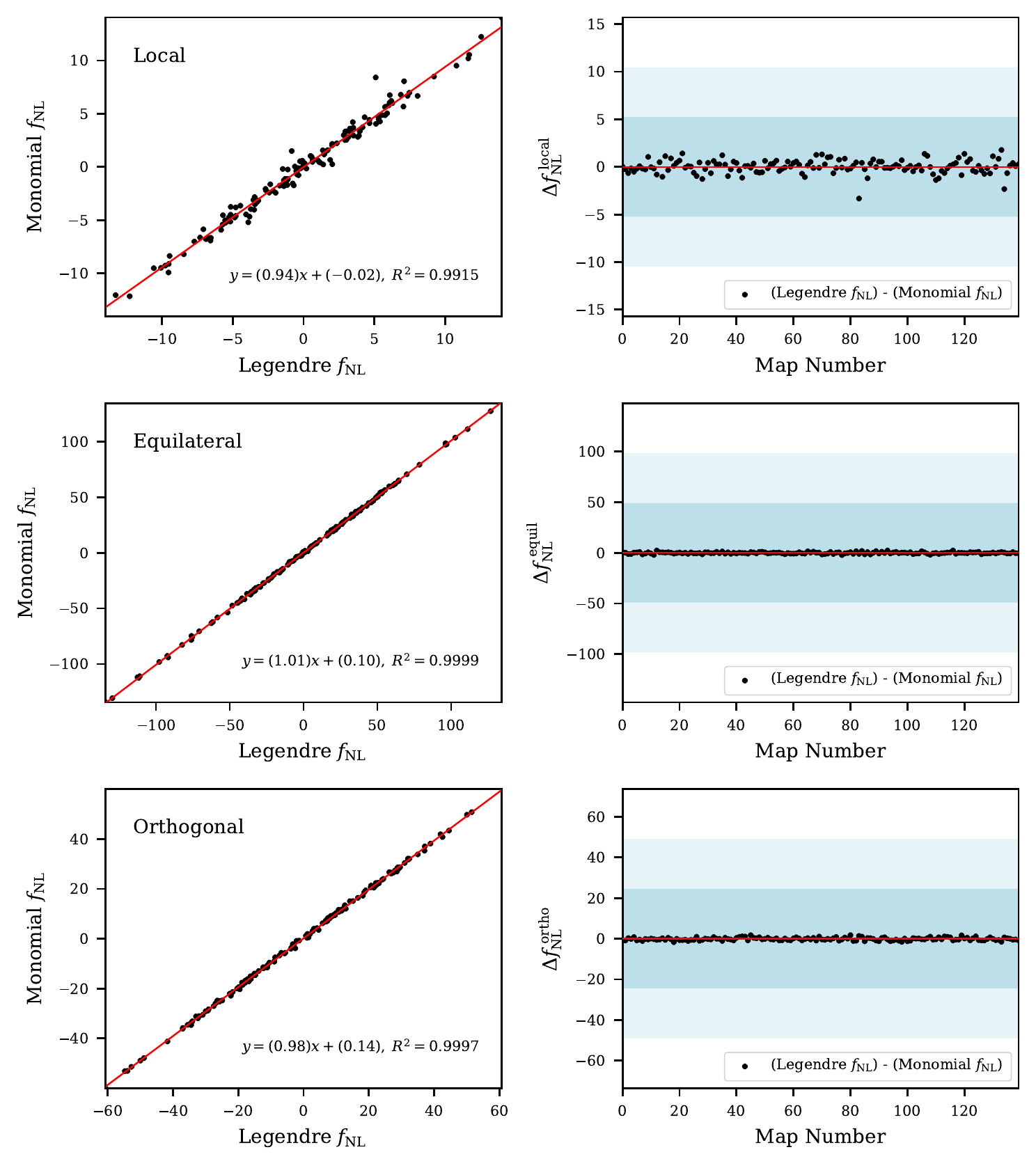}
	\caption{A map-by-map comparison of the $f_\text{NL}$ estimates evaluated using the Monomials (KSW) and Legendre basis sets for three standard templates. The same as Figure \ref{fig:map_by_map_Legendre_Monomial_T} but using both temperature and E-mode polarization.}
    \label{fig:map_by_map_Legendre_Monomial_TP}
\end{figure*}

We also perform a map-by-map comparison of the \best\ pipeline with Planck's Modal estimator. The results are shown in Figure \ref{fig:map_by_map_Legendre_Modal}. Overall, we find the two methods to be consistent and have $\fNL$ correlations varying between $0.93$ and $0.97$. The level of scatter shown here is within the bounds seen between the different estimators used in Planck \cite{PlanckCollaboration2013}. The local template shows a slightly higher level of discrepancy due to the limited $k$-range in the Legendre basis mentioned above. For a detailed analysis of the reason why different estimators are not perfectly correlated, we refer to the Appendix of \cite{PlanckCollaboration2013}. 

\begin{figure*}[htbp!] 
	\centering    
	\includegraphics{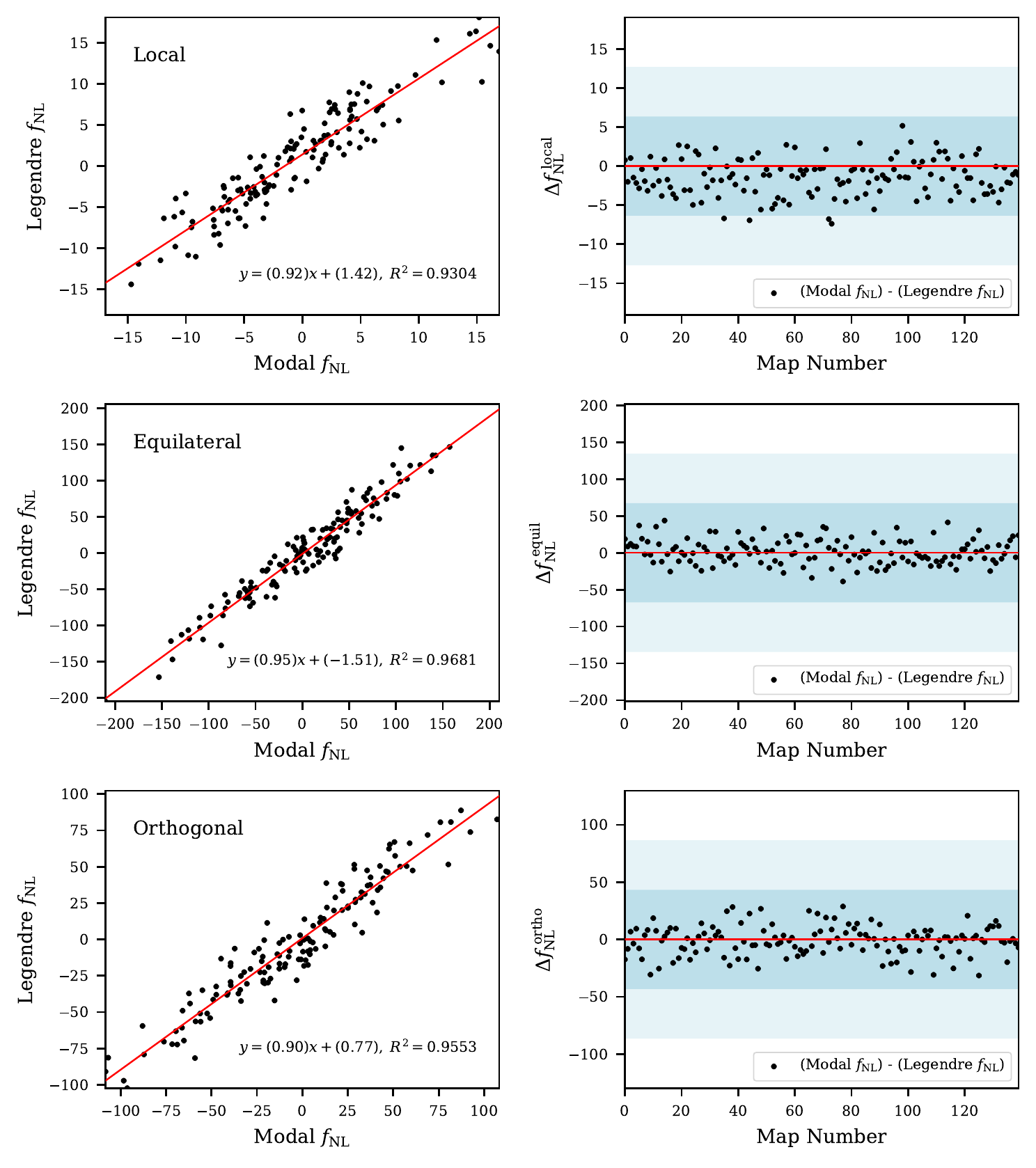}
	\caption{A map-by-map comparison of the $f_\text{NL}$ estimates obtained from the \best's Legendre basis set against the Modal estimator results of the Planck 2018 analysis \cite{PlanckCollaboration2018}. The first 160 FFP10 simulations are used here. On the left-hand side are scatter plots where each simulation is represented by a point according to $f_\text{NL}$ estimates of standard templates. Their linear best-fit lines are shown in red. Differences in the estimates from the two routines are shown map-by-map on the right-hand side, together with the $1\sigma$ and $2\sigma$ levels shaded in blue. Overall, \best\ and Modal are in good agreement without any significant systematic errors.}
    \label{fig:map_by_map_Legendre_Modal}
\end{figure*}

\subsection{Optimality of the bispectrum estimator}     \label{section:appendix_optimality}

In this section, we test the optimality of \best\ estimation to evaluate the validity of the Fisher matrix approximation for writing down the CMB bispectrum likelihood \eqref{eqn:bispectrum_likelihood_formal}.

The error given by the Fisher matrix in the likelihood is a rather theoretical quantity; it measures the expected scatter of the estimated $\fNL$ values across different realisations of the Universe. Unfortunately, we only have one universe to observe from, so simulated maps are used instead. As before, 160 CMB maps from FFP10 simulations after SMICA component separation \cite{PlanckCollaboration2015simulations,PlanckCollaboration2018hfi,PlanckCollaboration2018component} are plugged into \best\ pipeline to compute the $\fNL$ estimates. The simulations are based on Gaussian initial conditions.

The analysis would benefit from having more simulations, but the publicly available maps from the Planck Legacy Archive are only usable up to simulation number 160 \footnote{The files are named dx12\_v3\_smica\_cmb\_mc\_00xxx\_raw.fits, where xxx corresponds to the simulation number.}, since the ones after then are currently erroneous. Their temperature and polarisation maps yield angular power spectra that are consistent with the background cosmology, but the cross spectra ($ C^\mathrm{TE}_\ell $) are completely inconsistent.

Figure \ref{fig:triangle_consistency_T} and \ref{fig:triangle_consistency_TP} show comparisons between the distribution of the $\fNL$ estimates from simulations and the expected distribution computed using the Fisher matrix. The Python library \texttt{GetDist} \cite{Lewis2019getdist} was used to plot the sample points as contours. The irregular shapes of the contours are mainly due to the limited number of simulations used. Overall, we find that the Fisher matrix well approximates the error except for a slight underestimation at between 4-10\% level.

\begin{figure*}[htbp!]
	\centering    
	\includegraphics{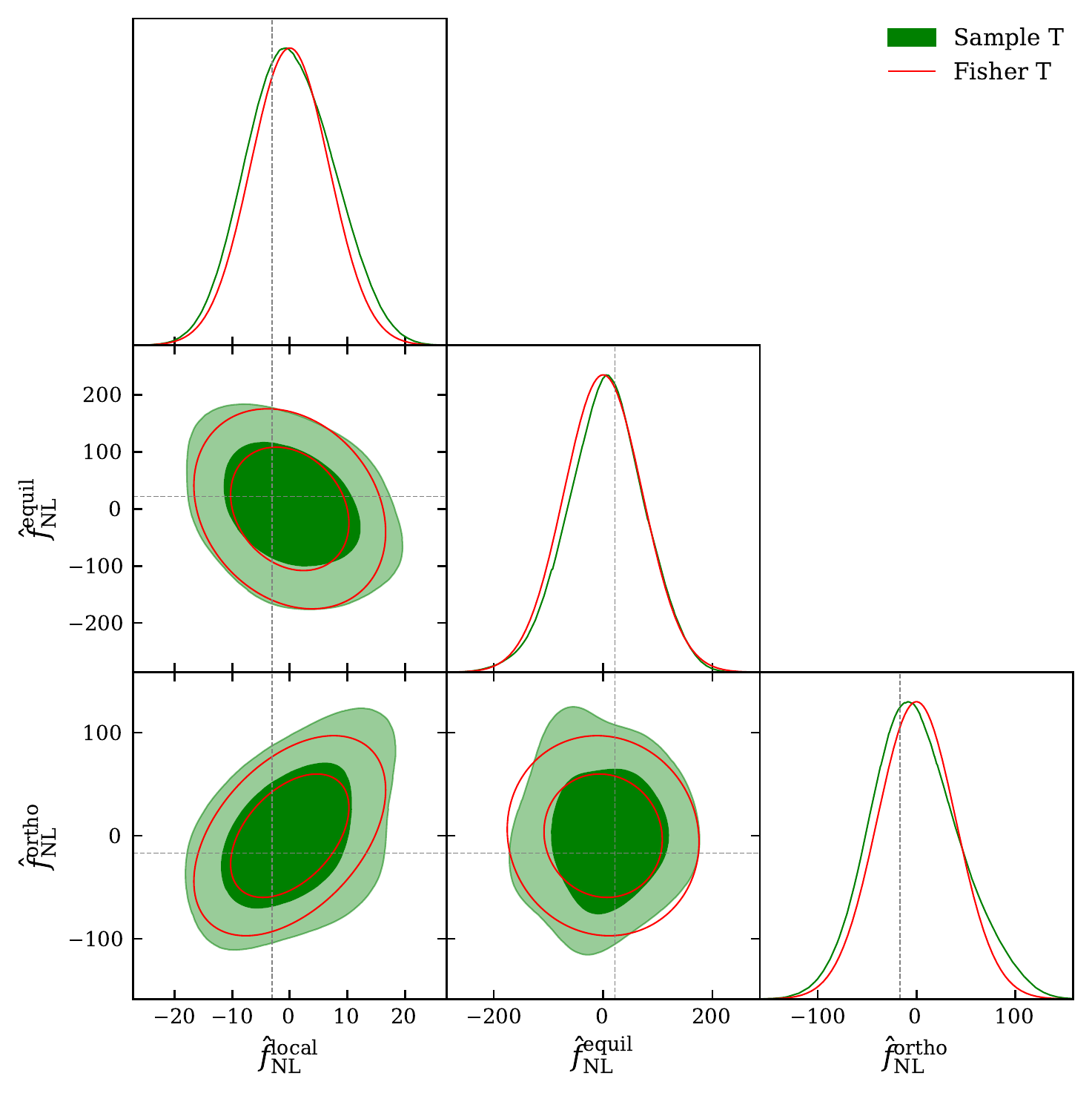}
	\caption{Comparison between the likelihoods estimated from the Fisher matrix and from $\fNL$ estimates of 160 FFP10 simulated temperature maps, under Gaussian initial conditions. The dashed lines indicate the marginal $\fNL$ estimates from observations. The two likelihoods are consistent with each other, even though the Fisher one underestimates the marginalized error by up to $10$\%. Note that the contour looks slightly irregular due to the limited number of samples.  }
	\label{fig:triangle_consistency_T}
\end{figure*}

\begin{figure*}[htbp!] 
	\centering    
	\includegraphics{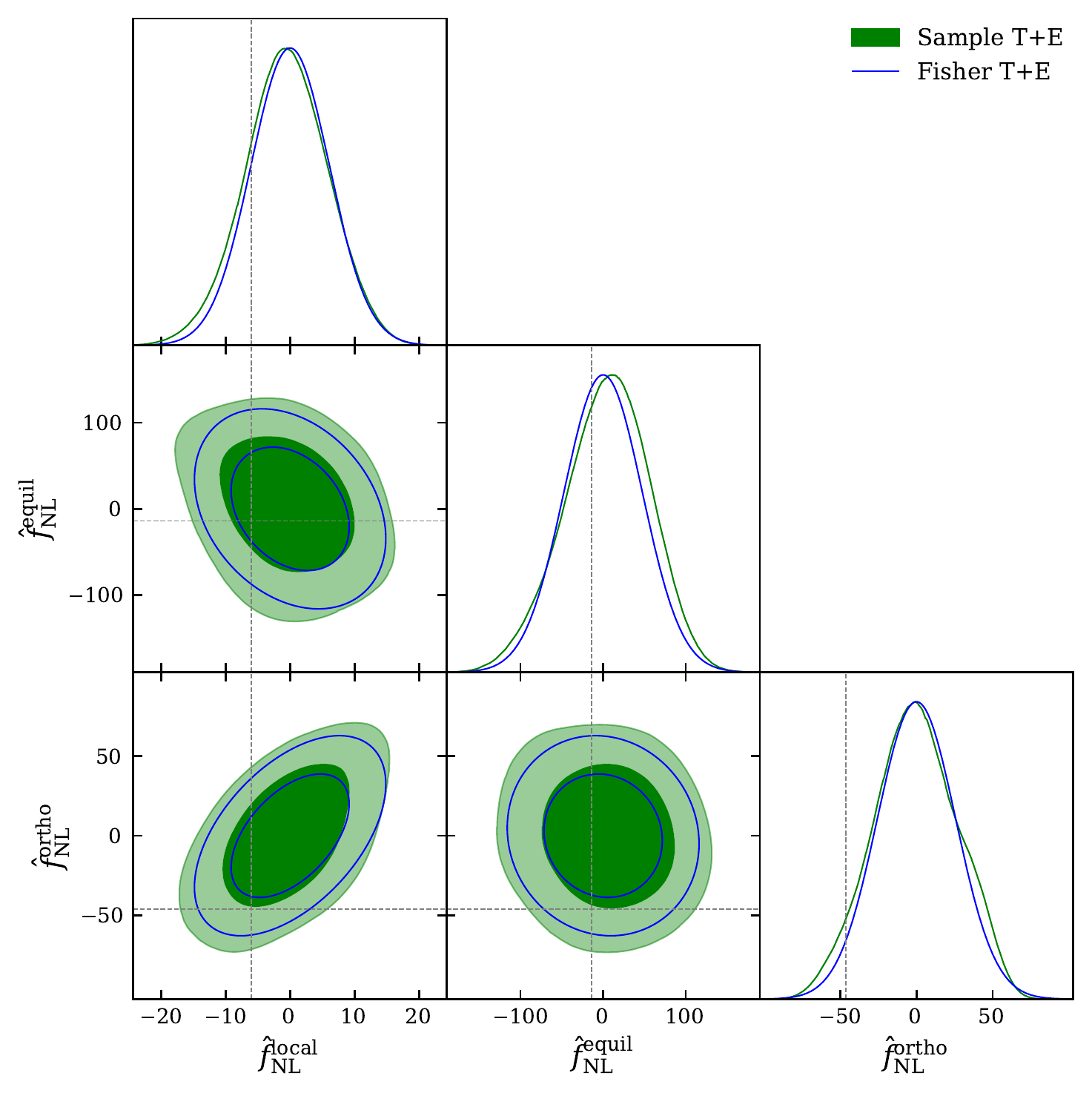}
	\caption{Comparison between the likelihoods estimated from the Fisher matrix and from $\fNL$ estimates of 160 FFP10 simulated temperature and polarization maps, under Gaussian initial conditions. Same as Figure \ref{fig:triangle_consistency_T} but with both temperature and polarization included. We again find that the two likelihoods are consistent with each other despite the underestimation of the marginalized error by up to $9$\%.  }
	\label{fig:triangle_consistency_TP}
\end{figure*}

\newpage

\bibliography{references}

\end{document}